\documentclass[journal]{IEEEtran}
\usepackage{amsmath,amsfonts}
\usepackage{algorithmic}
\usepackage{algorithm}
\usepackage{array}
\usepackage[caption=false,font=normalsize,labelfont=sf,textfont=sf]{subfig}
\usepackage{textcomp}
\usepackage{stfloats}
\usepackage{url}
\usepackage{verbatim}
\usepackage{graphicx}
\usepackage{cite}
\begin{document}

\title{Why Do Opinions and Actions Diverge? A Dynamic Framework to Explore the Impact of Subjective Norms}

\author{Chen Song, Vladimir Cvetkovic, Rong Su
        % <-this % stops a space
\thanks{This work was done as part of the joint PhD programme between Nanyang Technological University, Singapore and KTH Royal Institute of Technology, Stockholm, Sweden. This work was supported in part by the National Research Foundation, Singapore through its Medium Sized Center for Advanced Robotics Technology Innovation (CARTIN) under Project WP2.7. This work was also supported in part by the project `Humanizing the Sustainable Smart City (HiSS)' within the KTH Digital Futures research program. \textit{(Corresponding author: Vladimir Cvetkovic.)}}% <-this % stops a space
\thanks{Chen Song and Rong Su are with the School of Electrical and Electronic Engineering, Nanyang Technological University, Singapore 639798 (e-mail: song0249@e.ntu.edu.sg; rsu@ntu.edu.sg).}% <-this % stops a space
\thanks{Vladimir Cvetkovic is with the School of Architecture and the Built Environment, KTH Royal Institute of Technology, Stockholm 114 28, Sweden (e-mail: vdc@kth.se).}
%\thanks{.}
}

\markboth{Preprint submitted to IEEE Transactions on Computational Social Systems}%
%\markboth{IEEE Transactions on Computational Social Systems,~Vol.~14, No.~8, August~2021}%
{Song \MakeLowercase{\textit{et al.}}: Why Do Opinions and Actions Diverge? A Dynamic Framework to Explore the Impact of Subjective Norms}


\maketitle

\begin{abstract}
Socio-psychological studies have identified a common phenomenon where an individual's public actions do not necessarily coincide with their private opinions, yet most existing models fail to capture the dynamic interplay between these two aspects. To bridge this gap, we propose a novel agent-based modeling framework that integrates opinion dynamics with a decision-making mechanism. More precisely, our framework generalizes the classical Hegselmann-Krause model by combining it with a utility maximization problem. Preliminary results from our model demonstrate that the degree of opinion-action divergence within a population can be effectively controlled by adjusting two key parameters that reflect agents' personality traits, while the presence of social network amplifies the divergence. In addition, we study the social diffusion process by introducing a small number of committed agents into the model, and identify three key outcomes: adoption of innovation, rejection of innovation, and the enforcement of unpopular norms, consistent with findings in socio-psychological literature. The strong relevance of the results to real-world phenomena highlights our framework's potential for future applications in understanding and predicting complex social behaviors.
\end{abstract}


\section{Introduction} \label{section 1}
\IEEEPARstart{T}{he} study of opinion dynamics has drawn considerable interest from researchers across diverse disciplines, including social sciences, psychology, and engineering, particularly in the systems and control community \cite{ref1}. Opinion dynamics examines how individuals' opinions evolve through social interactions within a group. Each opinion dynamics model consists of three basic elements: the opinion expression format, the opinion dynamics environment, and the opinion fusion rule \cite{ref2}. Agents express their opinions in either continuous or discrete formats, and continuously update them through social interactions with others based on a predefined fusion rule \cite{ref2}. Some well-known continuous opinion dynamics have been proposed in the literature, including the Degroot model \cite{ref3}, the Friedkin-Johnsen model \cite{ref4}, and the bounded confidence opinion dynamics \cite{ref5, ref6}.

Among various models, the bounded confidence opinion dynamics (BCOD) have gained growing attention for integrating psychological factors into their opinion fusion rules. In BCOD, each agent is characterized by a confidence threshold that defines their confidence area, indicating the range of their acceptable opinions. In other words, each agent is only influenced by those whose opinions fall within their confidence area. This mechanism stems from the psychological concept of \textit{homophily} \cite{ref7}, where individuals tend to interact and connect with others sharing similar ideas. The most prominent instances of BCOD include the Deffuant-Weisbuch (DW) model \cite{ref5} and the Hegselmann-Krause (HK) model \cite{ref6}. A concise comparison of the two models, highlighting their distinct updating mechanisms and suitable application contexts, is provided in Appendix \ref{Appendix A}.  We also refer interested readers to \cite{ref1} and \cite{ref2} for a comprehensive review of opinion dynamics.  

\subsection{Motivations and Research Gap} \label{Section 1.1}

While existing opinion dynamics can effectively represent opinion formation processes, they fail to capture a commonly observed real-world phenomenon, where an individual's private opinions differ from their public actions. In socio-psychological literature, such discrepancies are often attributed to one's pressure to conform to the majority \cite{ref8}. Kuran \cite{ref9} highlighted the concept of `preference falsification', where individuals conceal their private preferences in public due to normative pressure, thus hindering the prediction of collective action such as social revolution. Asch's experiment \cite{ref10} demonstrated that individuals often conform to the unanimous pressure of a social group by making choices alighed with the majority, even when they blatantly contradict the fact. Prentice and Miller \cite{ref11} documented the existence of pluralistic ignorance, a psychological state in which one's private attitudes are different from others but they exhibit identical public behavior, in the context of students' attitudes toward alcohol consumption on campus. The abundant socio-psychological studies on conformity indicate the prevalence of opinion-action divergence, providing key motivations for our study. 

Before introducing the design concept of our proposed modeling framework, we first distinguish between the concepts of opinion and action (or decision). In this study, opinion is interpreted as an individual's internal view or judgment on a particular issue, regarded as private and known only to themselves. In contrast, action or decision is defined as the observable outcome of an individual's behavior on the same issue, which is considered public and visible to others. This distinction between opinion and action (or decision) serves as the fundamental assumption of our study and aligns well with principles established in socio-psychological literature. In Prentice and Miller \cite{ref11}, social norms are categorized as subjective norms defined by people's public behavior including all types of public statements, and actual norms that reflect their private views. When people's public behavior fails to accurately represent their private attitudes, the group's subjective norms will diverge from its actual norms, leading to opinion-action divergence \cite{ref11}. For the sake of clarity, the terms ``action'' and ``decision'' are used interchangeably in this study, as are the terms ``opinion'' and ``attitude''.  

Previous studies have primarily focused on either the opinion formation process through opinion dynamics or decision-making mechanisms using optimization methods and game theory \cite{ref12}. However, opinions and actions are closely interconnected in reality: an individual's opinions strongly influence their decisions, while the observed actions of others, in turn, play a crucial role in shaping their own opinions. To the best of our knowledge, few studies have explored the interplay between opinion and action by integrating opinion dynamics with a decision-making mechanism, highlighting a clear research gap. Zino et al. \cite{ref13} is one of the few works that developed a dynamic model for the coevolution of opinions and decisions, which serves as a key reference for our study. 

\subsection{Contributions of this Paper} \label{Section 1.2}

Inspired by these preliminary works, the first key contribution of our study is to address the research gap by developing a novel modeling framework that integrates opinion dynamics with a decision-making mechanism. Our proposed framework introduces notable improvements over prior research, particularly the framework developed by Zino et al. \cite{ref13}. To be more specific, it is characterized by three distinctive features that enhance its realism and applicability, setting it apart from existing models.

First, we employ an agent-based BCOD model, namely the Hegselmann-Krause (HK) model \cite{ref6}, to represent the opinion evolution process. As mentioned earlier, BCOD models are generally regarded as more intuitive and realistic as they are rooted in psychological principles, compared to the equation-based Friedkin-Johnsen model \cite{ref4} adopted by Zino et al \cite{ref13}. It is worth noting that our dynamic framework builds upon and extends the classical HK model \cite{ref6} to accomodate more complex interactions between opinion and action, with the HK model serving as a special case of our generalized model. As far as we are aware, no existing research has proposed a coevolution modeling framework utilizing agent-based BCOD models. The details of our model, including its generalization of the HK model, will be discussed further in Section~\ref{section 3}.   

Second, our framework incorporates a more realistic assumption that each agent's opinion remains private and visible only to themselves. Thus, each agent's opinion is influenced solely by the observed actions of others, rather than by others' private opinions. This assumption contrasts with the simplifications commonly employed in most existing opinion dynamics and Zino et al. \cite{ref13}, where opinions are treated as public and are capable of directly influencing one another. By accounting for the private nature of opinions, our framework captures more nuanced and realistic dynamics of social interactions. Moreover, we extend the concept of action from a discrete binary variable, as defined in Zino et al. \cite{ref13}, to a continuous variable, enabling the model to be applied in a more general context. A detailed interpretation of the opinion and action variables is provided in Section~\ref{section 3}. 

\begin{figure}[!t]
\centering
\includegraphics[width=2.5in]{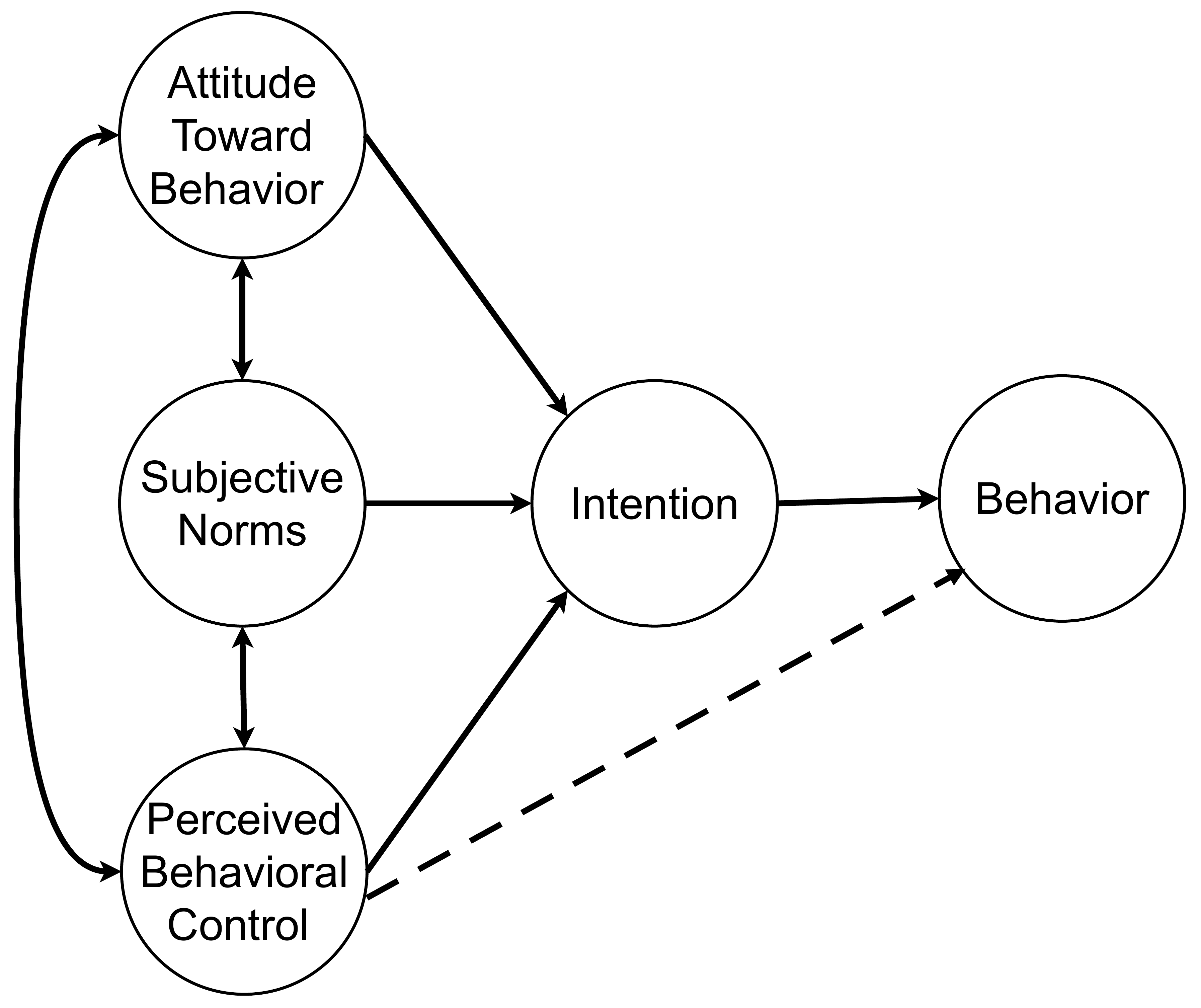}
\caption{Theory of Planned Behavior \cite{ref15}.}
\label{fig.1}
\end{figure}

Finally, our model features a sequential updating mechanism and integrates the concept of \textit{subjective norms} \cite{ref11,ref14}, as defined in social psychology, providing a framework more firmly rooted in established socio-psychological principles. More specifically, agents update their opinions and actions in two steps during each interaction. First, they revise their opinions under the influnence of observed actions of other agents. Then, they proceed to update their actions, guided by the group's subjective norms and their own updated opinions. This stands in contrast to Zino et al. \cite{ref13}, which considered a simultaneous update of opinions and actions, without taking into account the concept of subjective norms. According to the renowned Theory of Planned Behavior (TPB) \cite{ref15}, as shown in Fig.~\ref{fig.1}, individuals are assumed to exhibit rational behavior driven by their attitudes, subjective norms, and perceived control over their actions. Therefore, the enhancements in our framework provide a more realistic representation of opinion-action dynamics, supported by robust theoretical foundations.   

Building on our proposed modeling framework, we manage to observe and explain several real-world phenomena commonly observed in practice, including pluralistic ignorance \cite{ref11}, the adoption and rejection of innovations \cite{ref16}, and the enforcement of unpopular norms \cite{ref17}. The capability of our model to capture and examine a wide range of complex social dynamics constitutes the second major contribution of our study, demonstrating its strong relevance to the reality. This underscores that our study is not only deeply grounded in established socio-psychological theories but also has significant practical implications, providing potential interdisciplinary solutions to societal challenges.

The rest of this paper is organized as follows. Section~\ref{section 2} provides a comprehensive overview of the Hegselmann-Krause model, which serves as the basis for our framework. Section~\ref{section 3} elaborates on our model mechanism, including the structural components and mathematical formulations. Section~\ref{section 4} delivers preliminary results of our model, followed by a detailed discussion. Section~\ref{section 5} explores the relevance of our model for real-world social systems, offering insights into how future case studies could be designed to validate its predictive capabilities. Finally, Section~\ref{section 6} draws conclusions for our study and proposes directions for future research. 

\section{Preliminaries} \label{section 2}

In this section, we provide a detailed description of the Hegselmann-Krause (HK) model, including its underlying principles and mathematical formulations, which serves as the basis for our proposed model.

Consider a set of \( n \) agents, indexed by \( V = \{1, 2, \dots, n\} \). Each agent \( i \in V \) is characterized by an opinion variable \( x_i \in [0,1]\), which quantifies their attitudes toward a specific issue. For instance, in the context of a presidential election, \( x_i \) could represent one's political preferences, with \( x_i = 0 \) indicating a strong left-wing position, and \( x_i = 1 \) indicating a radical right-wing position. In addition, each agent is also assigned a confidence threshold variable \( \epsilon_i \in [0, 1] \), which defines their confidence area based on their opinion \( x_i \). For homogeneous HK model, where \( \epsilon_i = \epsilon, \, \forall \, i \in V \), each agent's confidence area corresponds to \( [x_i - \epsilon, x_i + \epsilon] \).

The HK model \cite{ref6} adopts the following opinion fusion rule: at each time step \( t \in \{0, 1, \dots, T\} \), each agent \( i \in V \) first identifies its neighbor set, \( N_i(t) \), which consists  of all agents whose opinions fall within its confidence area, given by \( N_i(t) = \{ j \in V \mid |x_i(t) - x_j(t)| \leq \epsilon \} \). The opinion of each agent \( i \) is then updated as:

\begin{equation} \label{e1}
x_i(t+1) = \frac{1}{|N_i(t)|} \sum_{j \in N_i(t)} x_j(t), \; \forall \, i \in V,
\end{equation}

\noindent where \( |N_i(t)| \) represents the number of agents in the neighbor set \( N_i(t) \). In other words, each agent updates its opinion to be the arithmetic mean of all opinions that fall within its confidence area, including its own opinion.

As reported in \cite{ref6}, the opinions of all agents, following the update rule (\ref{e1}), will eventually evolve into one of the three steady states: consensus, polarization, or fragmentation. The confidence threshold \( \epsilon \) plays a decisive role in shaping the steady state outcome: when \( \epsilon \) is very small, such as 0.1, the opinions at steady state are distributed across multiple distinct clusters, known as fragmentation. As \( \epsilon \) increases, the number of opinion clusters at steady state progressively decreases, ultimately resulting in consensus, where all agents reach an agreement and converge to a single shared opinion. 

The classical HK model assumes no constraints on agent interactions in terms of social network topology, implying that all agents form a fully connected network where every agent is connected with each other. However, the HK model can be extended to account for interactions within a specific social network topology. In such cases, the neighbor set of each agent is defined not only by the proximity of opinions but also by the underlying network connections. The neighbor set of agent \( i \) is now given by \( N_i(t) = \{ j \in V \mid |x_i(t) - x_j(t)| \leq \epsilon, \,  A_{ij} = 1 \} \), where \( A_{ij} = 1 \) indicates that agents \( i \) and \( j \) are connected in the network. By incorporating network topology into the HK model, this generalization facilitates the study of opinion dynamics within structured social networks. An overview of three basic network topologies commonly applied in the field of opinion dynamics, namely the complete graph \cite{ref18}, the small-world network \cite{ref19}, and the scale-free network \cite{ref20}, is presented in Appendix \ref{Appendix B}.

\section{Model formulation} \label{section 3}

In this section, we present a novel modeling framework that seamlessly integrates opinion dynamics with a decision-making mechanism. More precisely, this framework combines the HK model with a utility maximization problem, serving as a generalized extension of the classical HK model \cite{ref6}.

We consider a population of \( n \) agents, indexed by \( V = \{1, 2, \dots, n\} \). Within this framework, each agent \( i \) is characterized by two state variables: \( x_i \), representing the agent's opinion, as in the HK model, and \( y_i \), denoting the agent's action or decision regarding the same issue. Both opinion and action are continuous variables defined within the range [0,1], i.e., \( x_i, y_i \in [0,1] \). 

It is worth mentioning that the exact interpretations of \( x_i \) and \( y_i \) can vary across application contexts. However, from a generalized perspective, they represent the extent of an individual's attitudes or actions toward a particular issue, respectively. To provide a more intuitive explanation, we use the context of students' alcohol consumption behavior on campus, which is explored in Prentice and Miller \cite{ref11}, as an example for illustration. In this context, the opinion variable \( x_i \) could be interpreted as a student's attitude toward alcohol consumption, where a small value of \( x_i \) indicates strong opposition to drinking behavior, and a large \( x_i \)  reflects firm support for it. The action variable \( y_i \), on the other hand, could represent the student's actual alcohol consumption level, with a small \( y_i \) indicating no or minimal amount of alcohol intake, and a large \( y_i \) representing excessive drinking behavior. 

\begin{figure}[!t]
\centering
\includegraphics[width=2.5in]{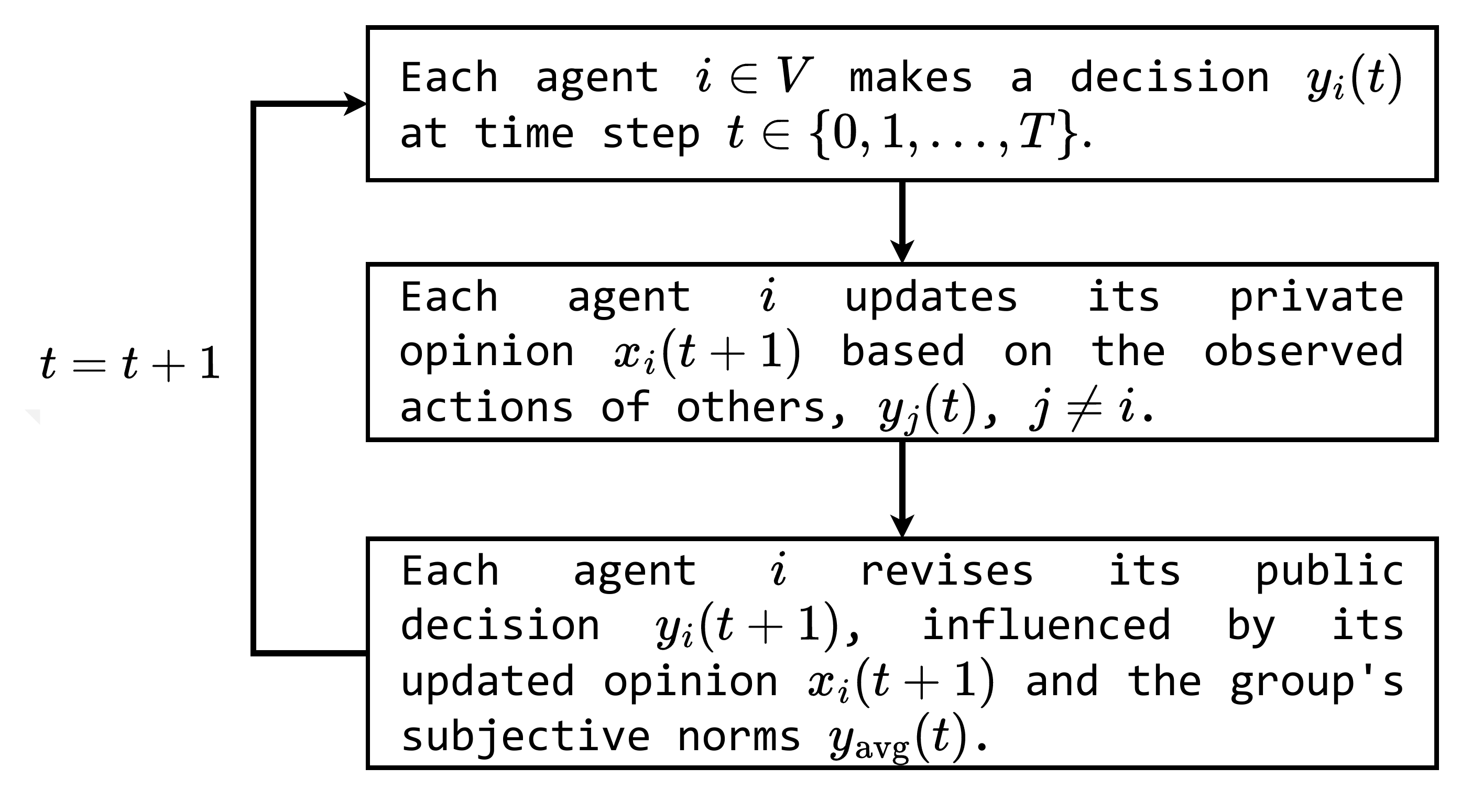}
\caption{Model flowchart.}
\label{fig.2}
\end{figure}

Our model employs a sequential updating mechanism, as illustrated in Fig.~\ref{fig.2}: at every time step, each agent first updates its opinion based on the observed actions of its neighbors, and then modifies its decision to maximize its utility function, which is constructed based on its updated opinion and the group's subjective norms. The details are elaborated in Section~\ref{section 3.1} and \ref{section 3.2}.

\subsection{Opinion Updating Mechanism} \label{section 3.1}

Our model initially assumes that all agents share a homogeneous confidence threshold, i.e., \( \epsilon_i = \epsilon, \, \forall \, i \in V \), and interact within a complete graph. It adopts the following opinion update rule: at every time step \( t \in \{0, 1, 2, \dots, T\}\), each agent \( i \in V \) determines its neighbor set  \( \mathcal{N}_i(t) \), defined as:

\begin{equation} \label{e2}
\mathcal{N}_i(t) = \{ j \in V \setminus \{i\} \mid |x_i(t) - y_j(t)| \leq \epsilon \}.
\end{equation}

The opinion of each agent \( i \) is then updated as: 

\begin{equation} \label{e3}
x_i(t+1) = \frac{\sum_{j \in \mathcal{N}_i(t)} y_j(t) + x_i(t)}{|\mathcal{N}_i(t)| + 1},
\end{equation}

\noindent where \( |\mathcal{N}_i(t)| \) denotes the number of agents in the neighbor set \( \mathcal{N}_i(t)\). 

Our model formulation extends the classical HK model by substituting the neighbors' opinions, \(x_j(t)\), with their actions, \(y_j(t)\). According to (\ref{e2}) and (\ref{e3}), agent \( i \)'s neighbors are now defined as those whose actions fall within its confidence area, and its opinion is updated to be the arithmetic mean of all its neighbors' actions and its own opinion. This modification is grounded in our realistic assumption that an agent's opinion is private and visible only to itself, whereas one's action is observable to others. Thus, each agent's opinion is influenced by the collective actions of its neighbors. 

In our modeling framework, the confidence threshold  \( \epsilon \) also plays a crucial role in shaping each agent's opinion. This parameter reflects an agent's openness to adjusting its own opinion based on the observed actions of others. A smaller \( \epsilon \) reduces the size of an agent's neighbor set, as observed from (\ref{e2}), and then results in an updated opinion that remains closer to its prior value \( x_i(t) \), as indicated by (\ref{e3}). On the contrary, a larger \( \epsilon \) enlarges the neighbor set and leads to an updated opinion that aligns more closely with the average action of the entire group, denoted as \( y_{\rm avg}(t) = \sum_{i \in V} y_i(t) / |V| \), which reflects the group's subjective norms \cite{ref15} at time step \(t\). As a result, the parameter \( \epsilon \) captures the agent's susceptibility to social influence in shaping its opinion, and we refer to it as \textit{openness} in the rest of the paper.  

\subsection{Action Updating Mechanism} \label{section 3.2}
After updating opinion from \(x_i(t)\) to \(x_i(t+1)\), each agent \( i \) proceeds to compute its updated action \(y_i(t+1)\), which is obtained by solving a utility maximization problem. The utility of agent \(i\) at time \( t \), denoted as \(U_i(y_i)\), is a function of the agent's potential action at time step \( t+1 \). A simple and intuitive approach to defining the utility is shown as follows:   

\begin{equation} \label{e4}
\begin{array}{l}
U_i(y_i) = -\phi_i \cdot (y_i - x_i(t+1))^2 \\
\quad \quad \quad \quad - (1-\phi_i) \cdot (y_i - y_{\mathrm{avg}}(t))^2,
\end{array}
\end{equation}

\noindent where \( x_i(t+1) \) refers to the agent's updated opinion computed from (\ref{e3}), and \( y_{\rm avg}(t) \) represents the group's subjective norms at time step \( t \). 

The updated action of each agent \(i\) at time step \( t+1 \) is then derived from: 

\begin{equation} \label{e5}
y_i(t+1) = \arg\max_{y_i} U_i(y_i).
\end{equation}

According to (\ref{e4}), each agent's utility function comprises two main components. The first term, \((y_i - x_i(t+1))^2\), quantifies the discrepancy between the agent's upcoming action and its up-to-date opinion. The second term, \((y_i - y_{\mathrm{avg}}(t))^2\), measures the divergence between the agent's upcoming action and the group's perceived social norms.

Intuitively, a smaller opinion-action discrepancy indicates greater consistency between the agent's private attitude and public behavior, thus resulting in higher utility due to increased personal satisfaction. Simiarly, a smaller divergence between the agent's action and the group's subjective norms implies less peer pressure experienced by the agent, also contributing to higher utility. Therefore, the coefficients preceding the two terms, \(-\phi_i\) and \(-(1-\phi_i)\), are both negative, representing the relative weights assigned to the two components by agent \( i \).  

Finally, parameter \( \phi_i \in [0,1]\) reflects the extent to which agent \(i\) is committed to its own opinion when making a decision, which we refer to as \textit{commitment}. A larger \(\phi_i\) assigns greater weight to the first term in utility, indicating that agent \(i\) puts more emphasis on its own opinion than the group's subjective norms. For instance, in the extreme case where \(\phi_i = 1\), the optimal solution to (\ref{e5}) becomes \(y_i(t+1) = x_i(t+1)\), signifying complete consistency between the agent's action and opinion. On the other hand, a smaller \(\phi_i\) gives more weight to the second term, suggesting that agent \(i\) is more significantly influenced by the group's subjective norms. In the extreme case where \(\phi_i = 0\), the optimal solution to (\ref{e5}) is \(y_i(t+1) = y_{\mathrm{avg}}(t)\), implying full conformity to the group's subjective norms. Our model initially assumes homogeneous \(\phi\) for all agents, i.e., \( \phi_i = \phi, \, \forall \, i \in V \). It is apparent that when \( \phi = 1 \), our modeling framework reverts to the classical HK model because in this case \( y_i = x_i, \, \forall \, i \in V \). As a consequence, our model is indeed a generalization of the classical HK model \cite{ref6}. 

\section{Results and discussion} \label{section 4}

This section presents the simulation results of our model. We begin by examining the impact of two key parameters, \(\epsilon\) (representing agents' openness) and \(\phi\) (representing agents' commitment), on the divergence between their opinions and actions. A sensitivity analysis is then conducted to systematically evaluate how variations in these parameters influence the population-level opinion-action divergence. Next, we investigate the effects of two common social network topologies, including the small-world network (SWN) and scale-free network (SFN), on the alignment between a group's collective opinion and action. Finally, we study the social diffusion process by introducing a subset of stubborn agents, known as innovators, who adhere strictly to an innovative norm with perfect opinion-action alignment, enabling us to observe and explain several real-world phenomena frequently reported in socio-psychological studies.

In the first three scenarios involving homogeneous agents, the model is initialized as follows: The number of agents is set to \( n = |V| = 300 \), and the simulation is run for a total of \( T = 50 \) time steps. Each agent's initial opinion is distributed as \( x_i(0) \sim U[0, 1], \, \forall \, i \in V \), where \( U[0, 1] \) denotes the uniform distribution on \([0, 1]\). The initial action of each agent is set to be equal to its initial opinion, i.e., \( y_i(0) = x_i(0), \, \forall \, i \in V \). In addition, all agents are assigned identical parameter values for openness and commitment, i.e., \( \epsilon_i = \epsilon \) and \( \phi_i = \phi, \, \forall \, i \in V \).

In contrast to the homogeneous setup, the last scenario aims to explore the impact of a committed minority group on social diffusion by introducing 60 innovators into the model, comprising 20\% of the whole population. The subset of innovators, denoted as \( S \subset V \), is defined by distinct characteristics: each innovator \( s \in S \) is endowed with an openness \( \epsilon_s = 0 \), a commitment \( \phi_s = 1 \), and an initial opinion \( x_s(0) = 1 \), ensuring that \( y_s(t) = x_s(t) = 1, \, \forall \;  t \in \{0, 1, \dots, T\} \). In other words, these innovators remain unwaveringly committed to the innovative norm, represented by \( 1 \), in both opinions and actions throughout the simulation. For the remaining non-stubborn agents (\( i \in V \setminus S \)), their initial opinions are sampled from \( x_i(0) \sim U[0, 0.5] \), to ensure that their existing norms are different from the innovative norm. Moreover, their openness and commitment parameters are now drawn from uniform distributions over specified intervals rather than being assigned fixed, homogeneous values as in previous scenarios.  

\subsection{Impact of Openness (\(\epsilon\)) and Commitment (\(\phi\))} \label{section 4.1}

\begin{figure*}[!t]
\centering
\subfloat[]{\includegraphics[width=2.0in]{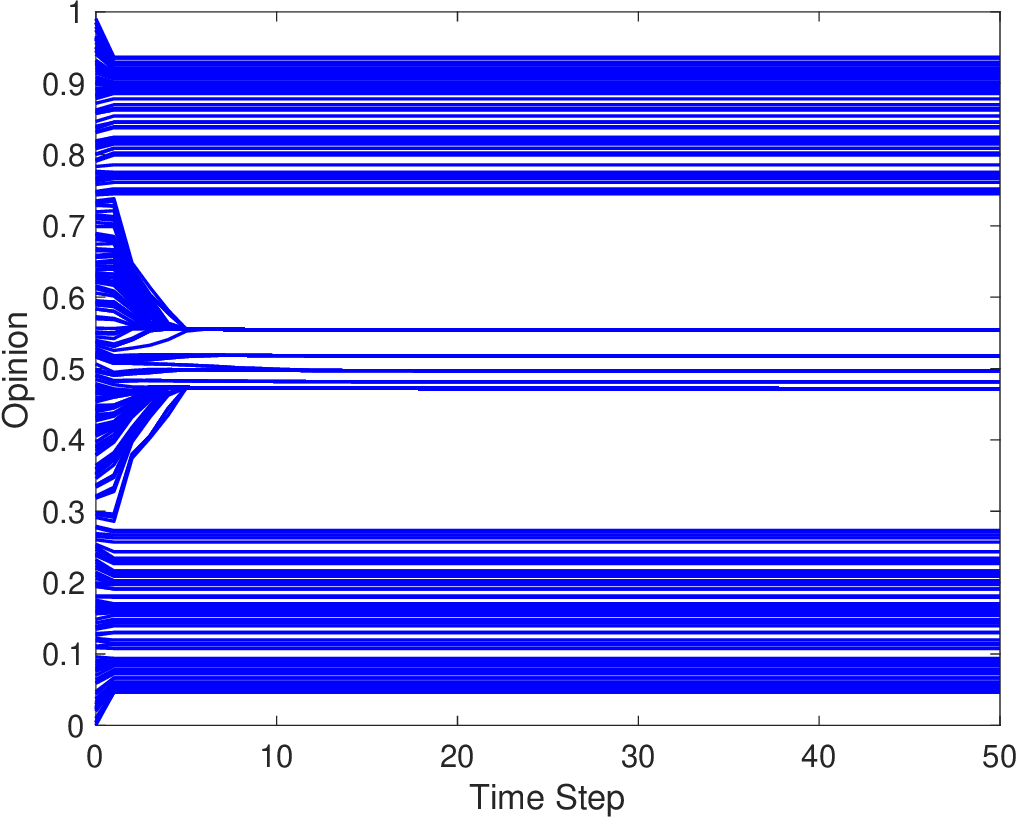}%
\label{fig3.1}}
\hfil
\subfloat[]{\includegraphics[width=2.0in]{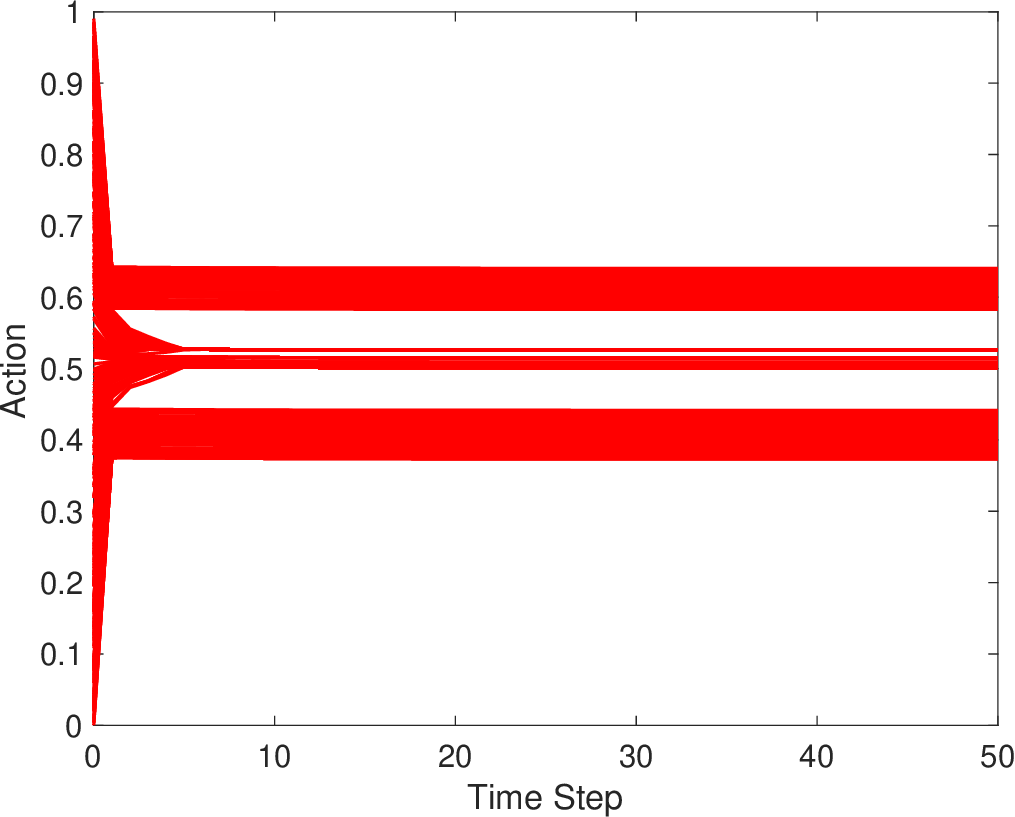}%
\label{fig3.2}}
\hfil
\subfloat[]{\includegraphics[width=2.0in]{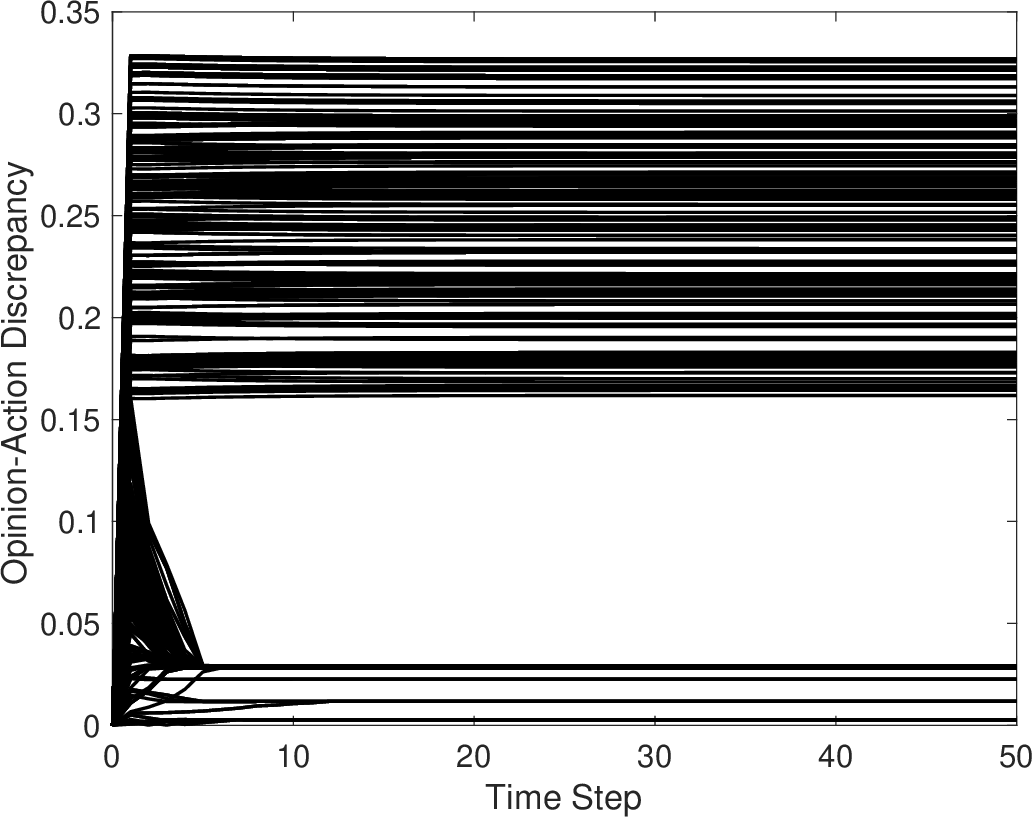}%
\label{fig3.3}}
\caption{Social dynamics for \(\epsilon=0.1\) and \(\phi=0.3\). (a) Opinion. (b) Action. (c) Opinion-action discrepancy.}
\label{fig3}
\end{figure*}

\begin{figure*}[!t]
\centering
\subfloat[]{\includegraphics[width=2.0in]{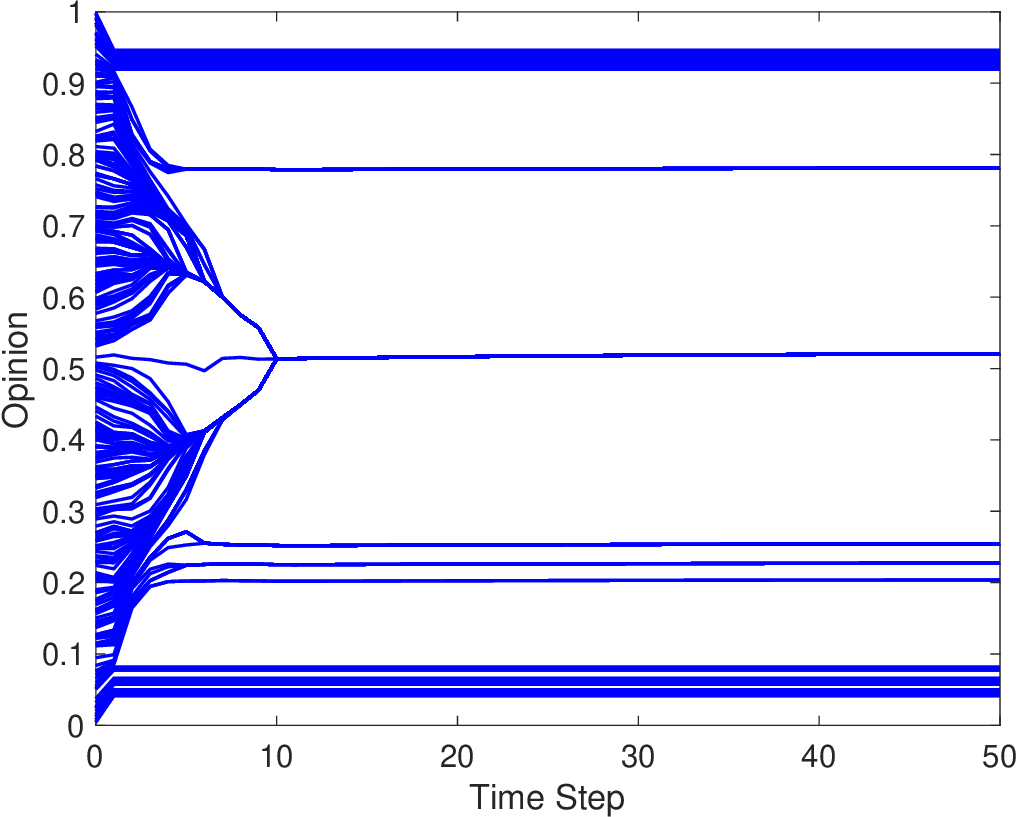}%
\label{fig4.1}}
\hfil
\subfloat[]{\includegraphics[width=2.0in]{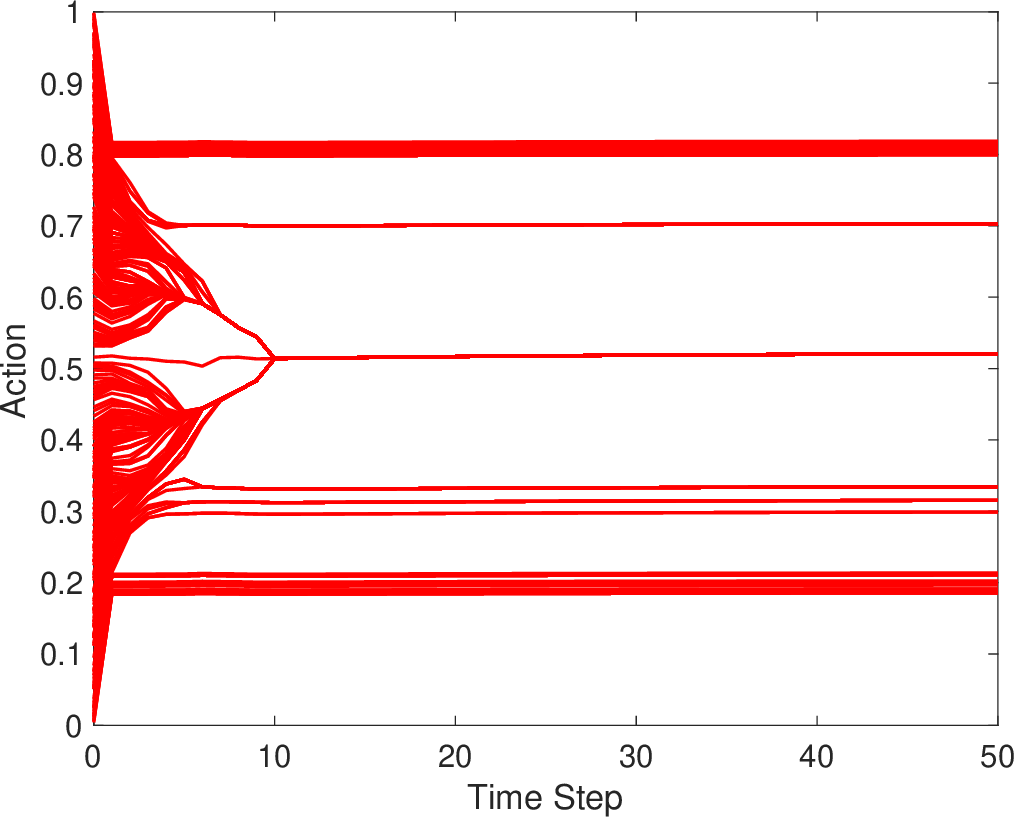}%
\label{fig4.2}}
\hfil
\subfloat[]{\includegraphics[width=2.0in]{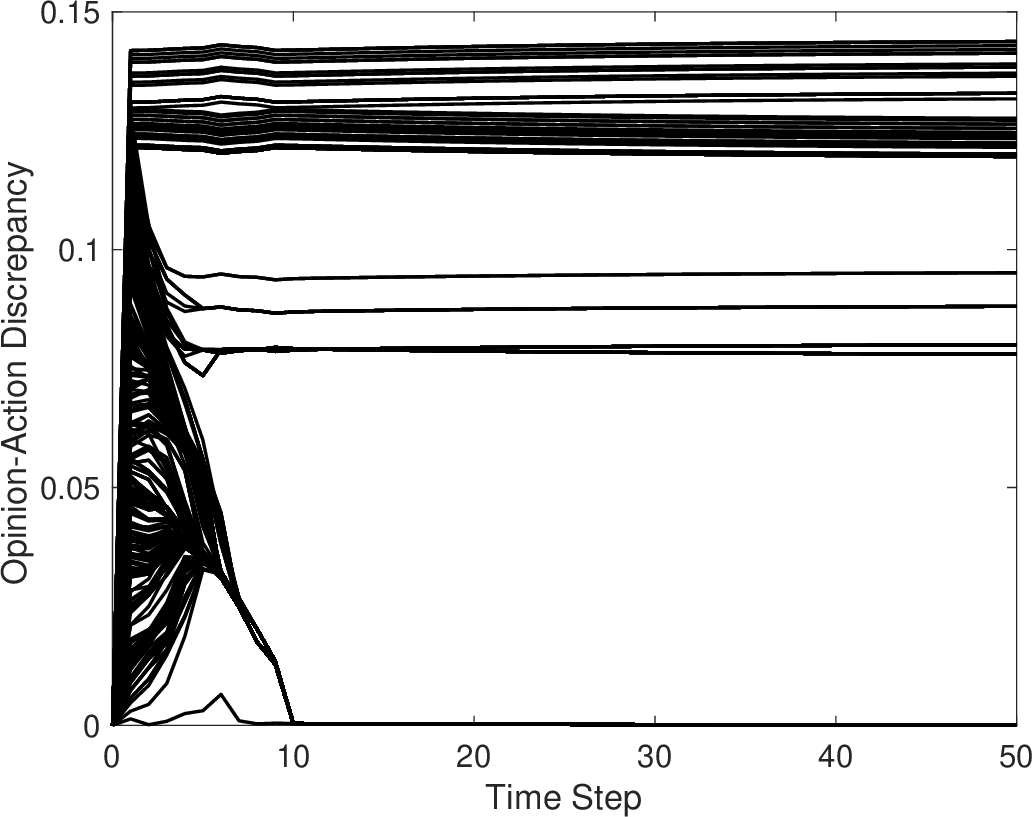}%
\label{fig4.3}}
\caption{Social dynamics for \(\epsilon=0.1\) and \(\phi=0.7\). (a) Opinion. (b) Action. (c) Opinion-action discrepancy.}
\label{fig4}
\end{figure*}

\begin{figure*}[!t]
\centering
\subfloat[]{\includegraphics[width=2.0in]{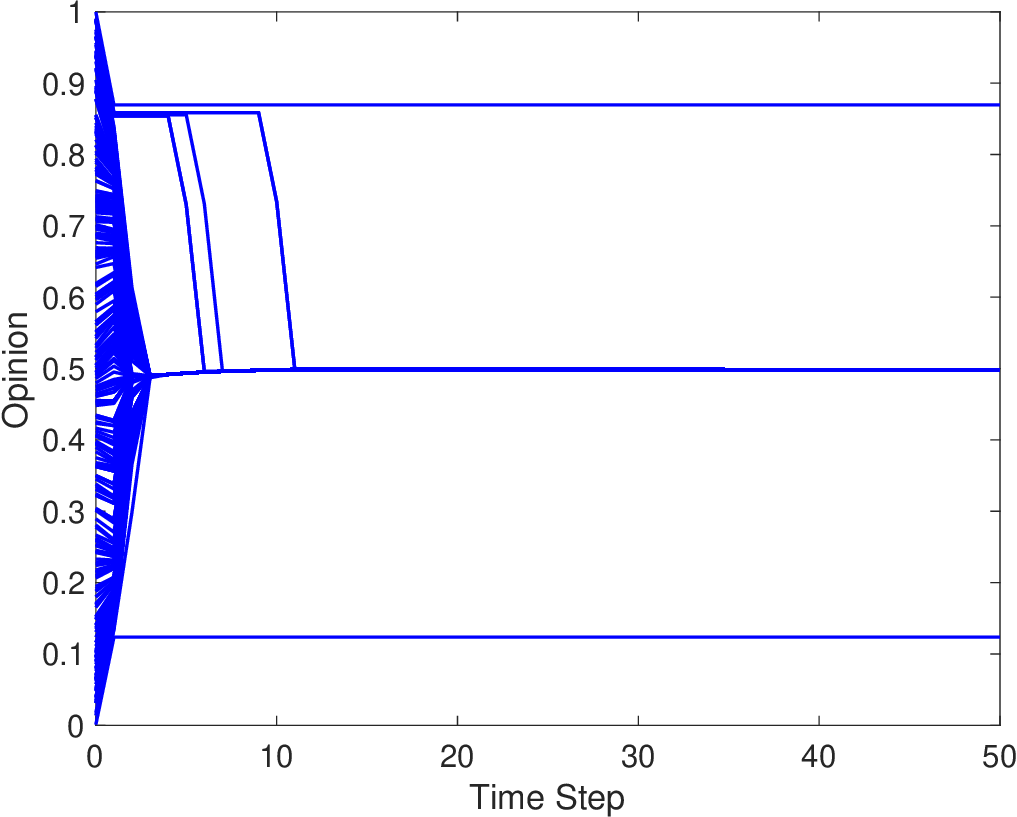}%
\label{fig5.1}}
\hfil
\subfloat[]{\includegraphics[width=2.0in]{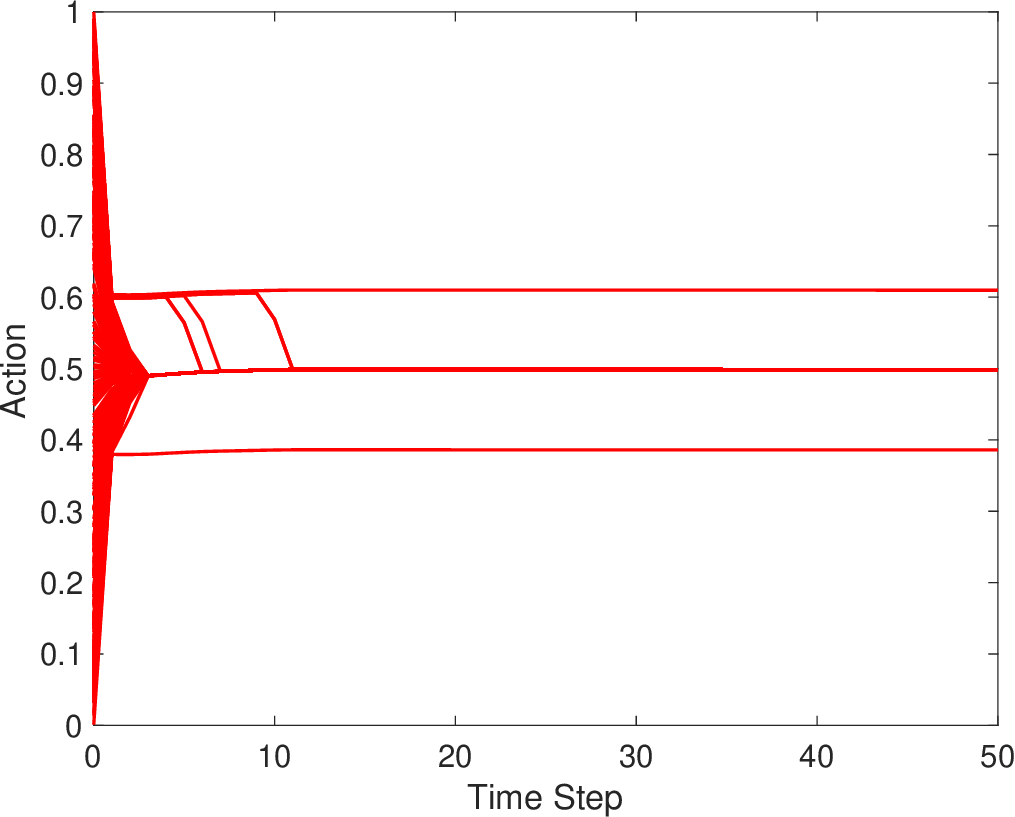}%
\label{fig5.2}}
\hfil
\subfloat[]{\includegraphics[width=2.0in]{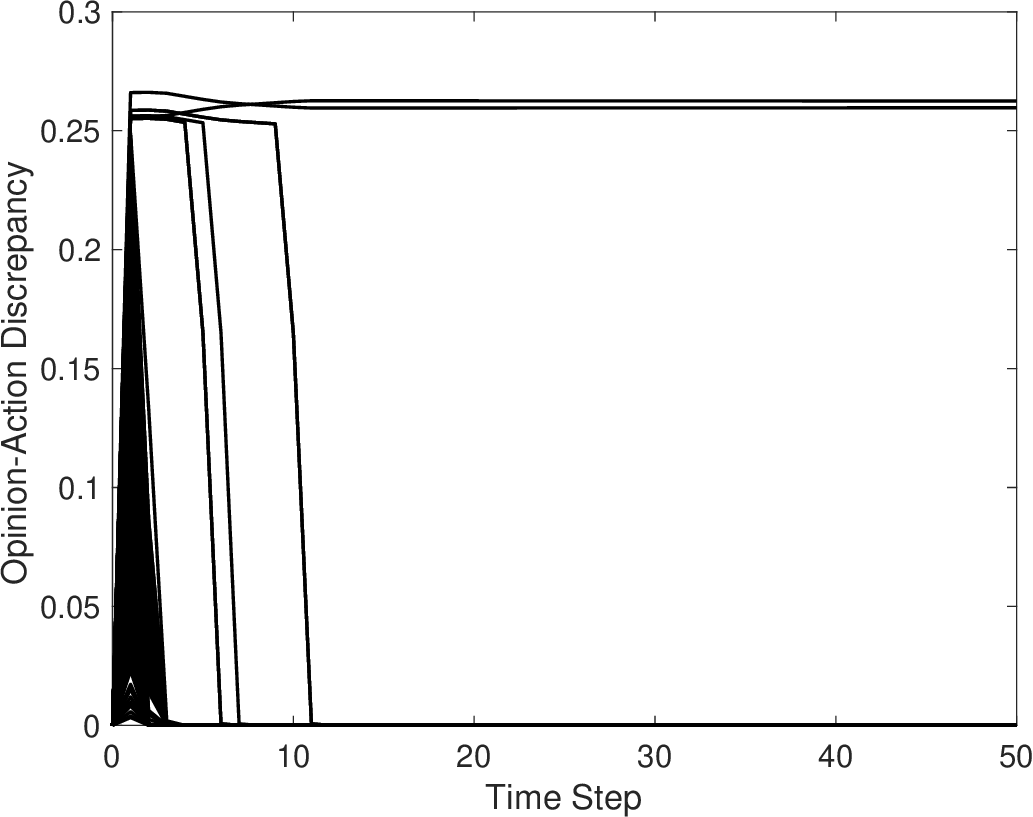}%
\label{fig5.3}}
\caption{Social dynamics for \(\epsilon=0.25\) and \(\phi=0.3\). (a) Opinion. (b) Action. (c) Opinion-action discrepancy.}
\label{fig5}
\end{figure*}

\begin{figure*}[!t]
\centering
\subfloat[]{\includegraphics[width=2.0in]{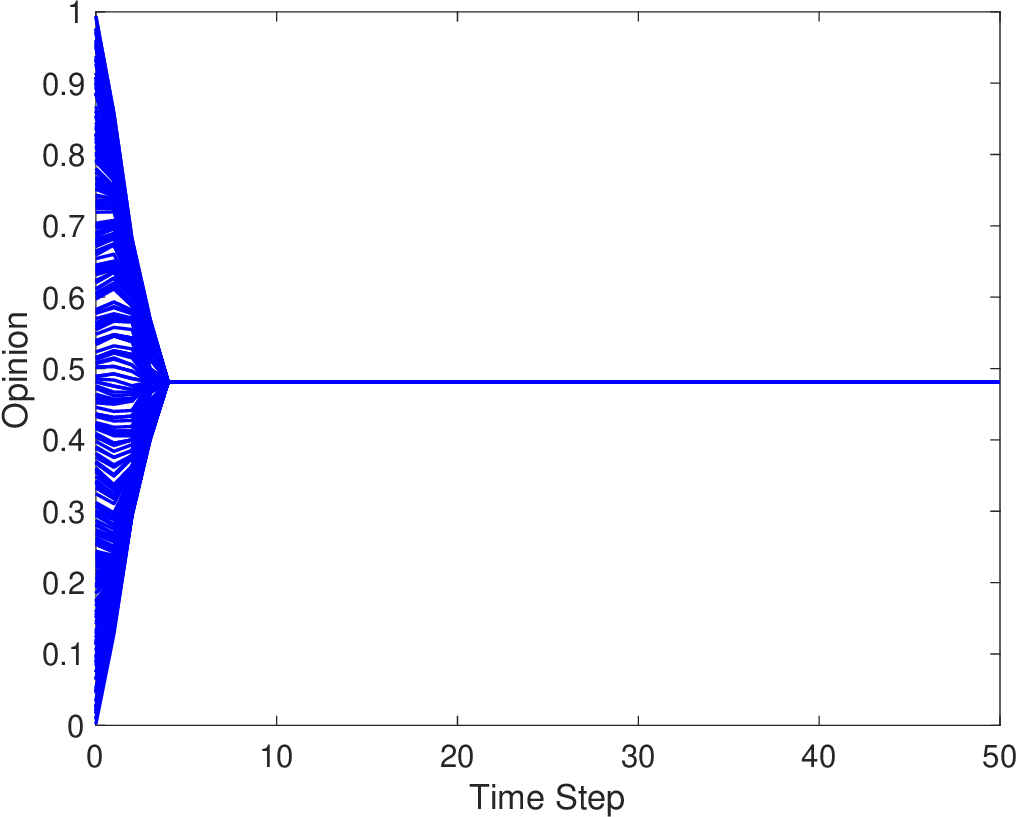}%
\label{fig6.1}}
\hfil
\subfloat[]{\includegraphics[width=2.0in]{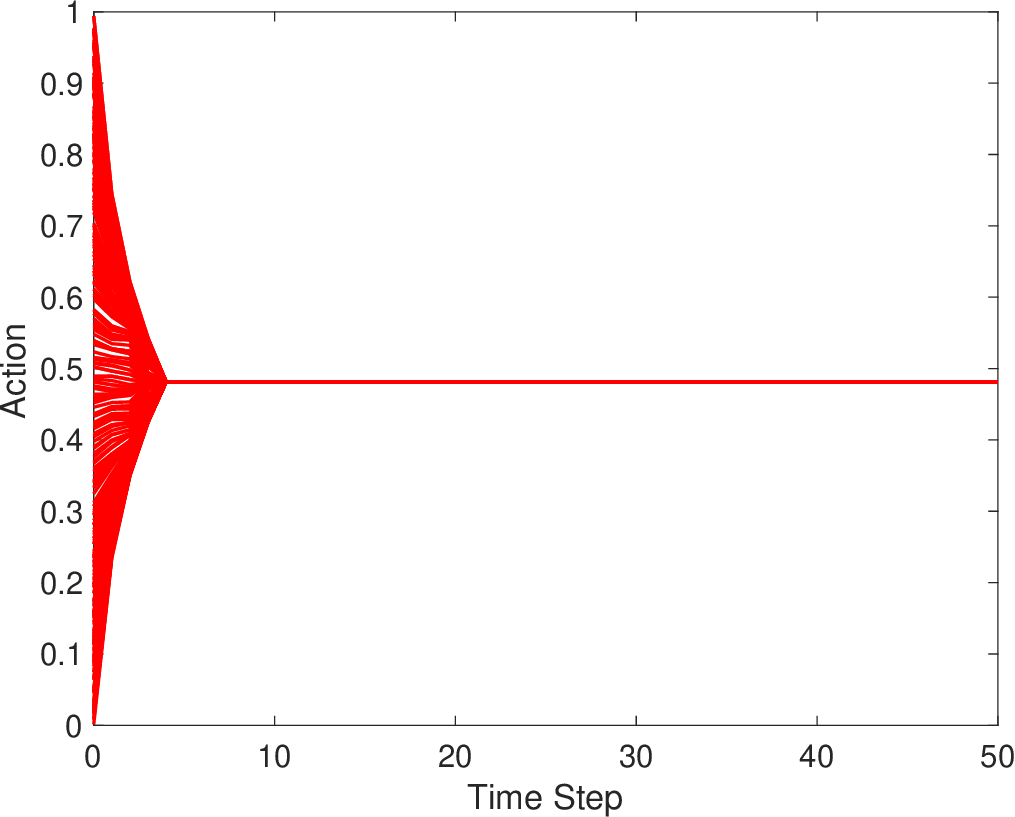}%
\label{fig6.2}}
\hfil
\subfloat[]{\includegraphics[width=2.0in]{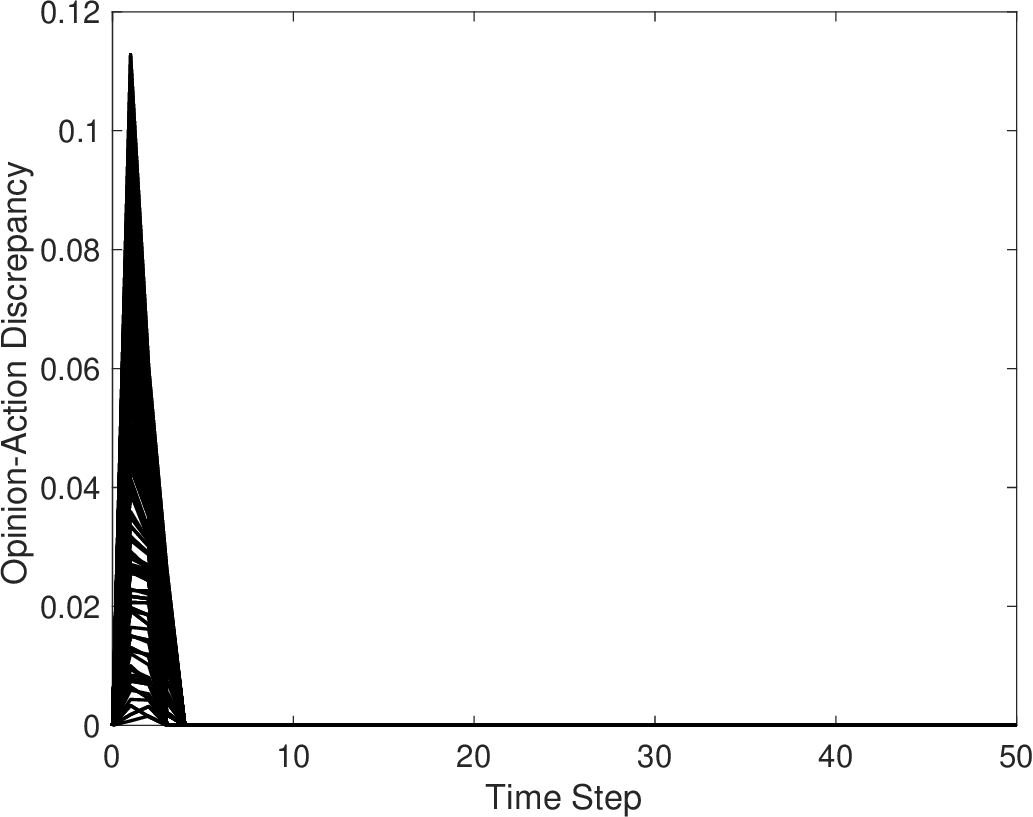}%
\label{fig6.3}}
\caption{Social dynamics for \(\epsilon=0.25\) and \(\phi=0.7\). (a) Opinion. (b) Action. (c) Opinion-action discrepancy.}
\label{fig6}
\end{figure*}

First, we qualitatively analyze the impact of openness (\( \epsilon \)) and commitement (\( \phi \)) on the opinion-action divergence among agents. In particular, we evaluate the outcomes of a cross-combination of parameter values: \(\epsilon = \{0.1, 0.25\}\) and \(\phi = \{0.3, 0.7\}\). The results are shown from Fig.~\ref{fig3} to Fig.~\ref{fig6}, which correspond to four parameter combinations. Each figure consists of three subplots: (a) opinion dynamics, (b) action dynamics, and (c) opinion-action discrepancy over time. The opinion-action discrepancy for each agent \( i \) at time step \( t \) is defined as: \( d_i(t) = |x_i(t) - y_i(t)|, \, \forall \, i \in V,  \, t \in \{0, 1, \dots, T\} \).

As shown in Fig.~\ref{fig3}, the results for the parameter combination (\(\epsilon = 0.1, \phi = 0.3\)) indicate a pronounced discrepancy between agents' opinions and actions. Specifically, it can be observed from Fig.~\ref{fig3}(a) and \ref{fig3}(b) that while the final opinion distribution spans most of the range from 0 to 1, the action profile is tightly concentrated around the center. In addition, Fig.~\ref{fig3}(c) reveals that most agents exhibit large opinion-action differences at steady state, which can reach up to 0.33. 

This phenomenon closely resembles the concept of pluralistic ignorance \cite{ref11}, where individuals hold diverse private attitudes yet display similar or even identical public behavior. In other words, when both \(\epsilon\) (openness) and \(\phi\) (commitment) are low, a significant opinion-action discrepancy arises within the group, potentially leading to pluralistic ignorance. The underlying explanation for this phenomenon is intuitively clear: when agents are close-minded and resist altering their opinions (low \(\epsilon\)), yet fail to act in alignment with their private beliefs (low \(\phi\)) due to peer pressure, they will form diverse opinions distinct from the conforming behavior. In essence, close-mindedness and weak alignment between opinions and actions contribute to the emergence of evident opinion-action divergence. 

For the parameter combination (\(\epsilon = 0.1, \phi = 0.7\)), as shown in Fig.~\ref{fig4}, the group's overall opinion-action discrepancy decreases significantly compared to Fig.~\ref{fig3}. The opinion and action distributions become closer to each other both in shape and cluster location, and the opinion-action differences for individual agents also drop below 0.15. 

The results of this scenario reveal another phenomenon: even though opinions and actions are generally consistent for most agents, the group naturally divides into multiple clusters with varied beliefs and behaviors. Such fragmentation within a group conceptually aligns well with real-world scenarios where a population splits into several smaller, cohesive subgroups exhibiting distinct habits, lifestyles, or cultural traits, such as the formation of communities \cite{ref21} and cultural identity \cite{ref22}. The reason behind this phenomenon is straightforward: when agents are reluctant to adjust their opinions (low \(\epsilon\)) and strongly adhere to their private attitudes when making a decision (high \(\phi\)), their opinion-action divergence becomes smaller, leading to diverse opinion clusters at steady state.   

In the third scenario (\(\epsilon = 0.25, \phi = 0.3\)), as illustrated in Fig.~\ref{fig5}, the opinion-action consistency among agents is also significantly improved compared to Fig.~\ref{fig3}. However, only three opinion or action clusters emerge at steady state, indicating a more focused distribution in contrast to Fig.~\ref{fig4}. In the last scenario as shown in Fig.~\ref{fig6}, where \(\epsilon = 0.25\) and \(\phi = 0.7\), all agents demonstrate perfect alignment between their opinions and actions. Moreover, the group achieves full consensus, with all agents converging to a single shared opinion and action.      

It can be observed from the last two scenarios' results that as each agent's openness (\( \epsilon \)) increases, their opinion-action discrepancy is greatly reduced, which can be attributed to the role that openness (\(\epsilon\)) plays in shaping their opinion. As indicated in (\ref{e2}) and (\ref{e3}), a larger \(\epsilon\) drives the agent's opinion to converge toward the group's subjective norms, represented by \(y_{\mathrm{avg}}(t)\). When openness (\(\epsilon\)) is high, regardless of whether  commitment (\(\phi\)) is low or high, opinions and actions are both predominantly influenced by the group's perceived social norms, making the group more likely to reach a consensus that reflects its subjective norms. 

This phenomenon is analogus to the conformity behavior observed in classical socio-psychological studies, such as the Asch's experiment \cite{ref10}. In the experiment, a single participant was placed among a unanimous majority whose common decision blatantly contradicted observable fact and common sense. Participants who gave in to the majority's incorrect decision, referred to as `yielding subjects' \cite{ref10}, reported that they had experienced overwhelming pressure during the experiment, which impaired their ability to reason effectively. In other words, the yielding subjects even lost the capacity to distinguish between the ground truth and intentionally misleading choices. Under these circumstances, their opinions and actions are both guided by the group's subjective norms due to a heightened openness to external opinions. This highlights a clear instance where conformity to the majority arises from an increased acceptance of imposed subjective norms, even when those norms are evidently incorrect.   

\subsection{Sensitivity Analysis of Openness (\(\epsilon\)) and Commitment (\(\phi\))} \label{section 4.2}

\begin{figure}[!t]
\centering
\includegraphics[width=2.5in]{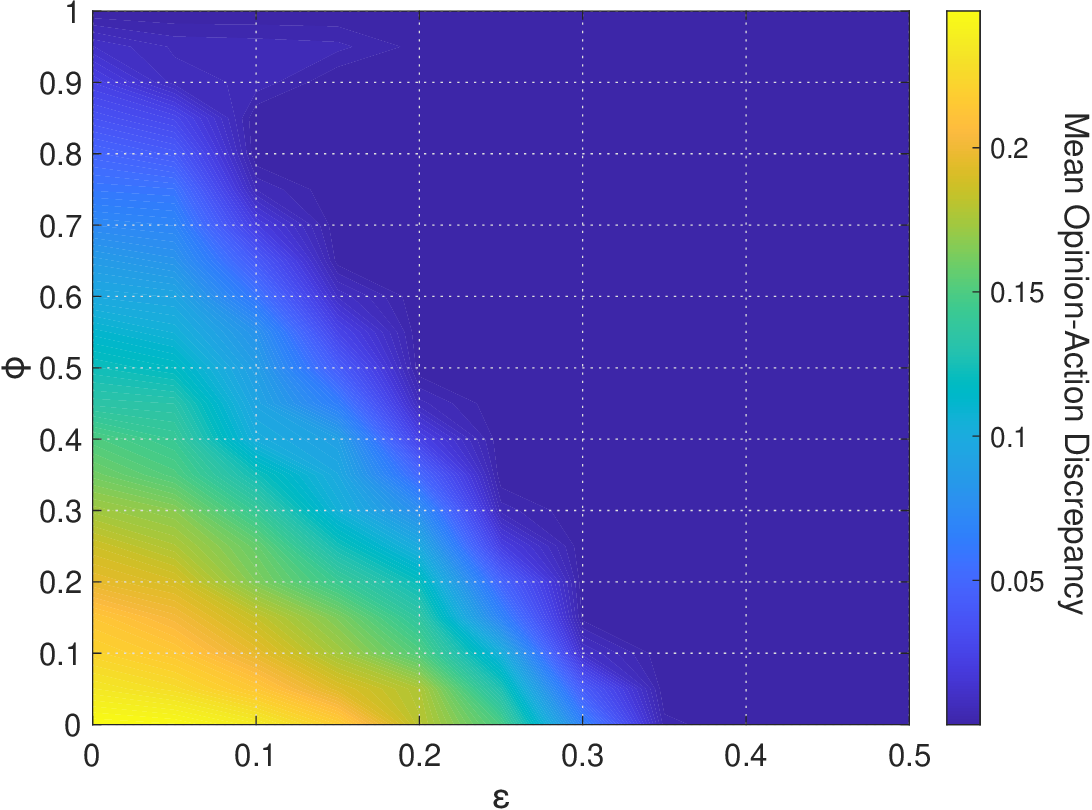}
\caption{Sensitivity analysis of \(\epsilon\) and \(\phi\).}
\label{fig7}
\end{figure}

From the results in Section~\ref{section 4.1}, it can be observed that as the sum of \(\epsilon\) and \(\phi\) increases, the group's aggregate opinion and action profiles become closer to each other. To explore the impact of these two parameters on the group-level opinion-action divergence in greater detail, we perform a sensitivity analysis of \(\epsilon\) and \(\phi\) for our model. The model is executed for all combinations of \(\epsilon \in [0, 0.5]\) and \(\phi \in [0, 1]\), with a step size of 0.05. The range of \(\epsilon\) is restricted to [0, 0.5] because according to the classical HK model, when \(\epsilon > 0.5\), the system is guaranteed to reach a consensus. The group-level opinion-action discrepancy is quantified as the average opinion-action discrepancy across all agents in the population at steady state, calculated as: 

\begin{equation} \label{e6}
D = \frac{1}{|V|}\sum_{i \in V} d_i^* = \frac{1}{|V|} \sum_{i \in V} |x_i^* - y_i^*|.
\end{equation}

\noindent where \(x_i^*\), \(y_i^*\), and \(d_i^*\) represent the steady-state opinion, action, and opinion-action discrepancy of agent \(i\), respectively. For each parameter combination, we perform 10 independent simulation runs and compute the average \(D\) value to account for stochastic variations introduced during the initialization process. The results of the sensitivity analysis are depicted in the contour plot, as shown in Fig.~\ref{fig7}.

The results in Fig.~\ref{fig7} confirm our observation that when the sum of \(\epsilon\) and \(\phi\) exceeds a critical threshold, every agent's opinion and action converge to a complete agreement, resulting in zero mean discrepancy for the entire group. This perfect alignment is approximately achieved when \(10\epsilon + 3\phi \geq 3.5\), as represented by the upper-right blue region of the plot. Conversely, when the (\(\epsilon\), \(\phi\)) combination falls below this boundary line, the average opinion-action discrepancy becomes positive, indicating that divergence between opinions and actions exists among certain agents in the population. These findings demonstrate that the magnitude of opinion-action divergence within a population can be effectively controlled by tuning their openness (\(\epsilon\)) and commitment (\(\phi\)) parameters.   

\subsection{Impact of Social Network Topology} \label{section 4.3}

\begin{figure*}[!t]
\centering
\subfloat[]{\includegraphics[width=2.0in]{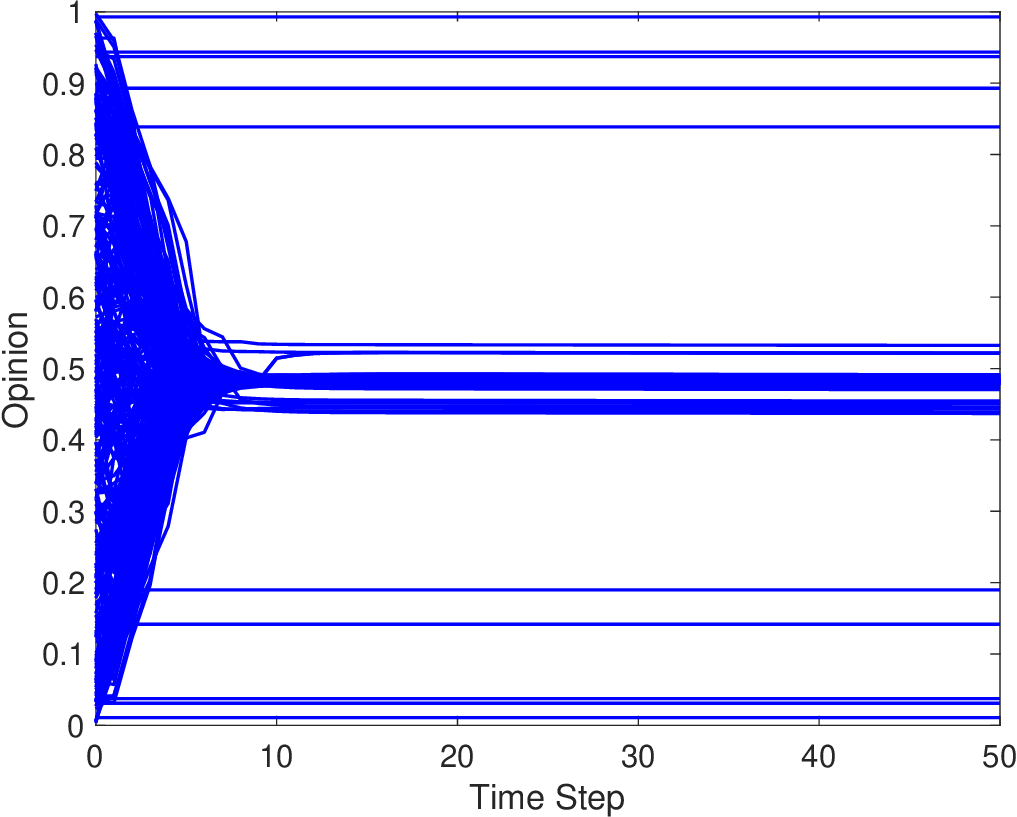}%
\label{fig8.1}}
\hfil
\subfloat[]{\includegraphics[width=2.0in]{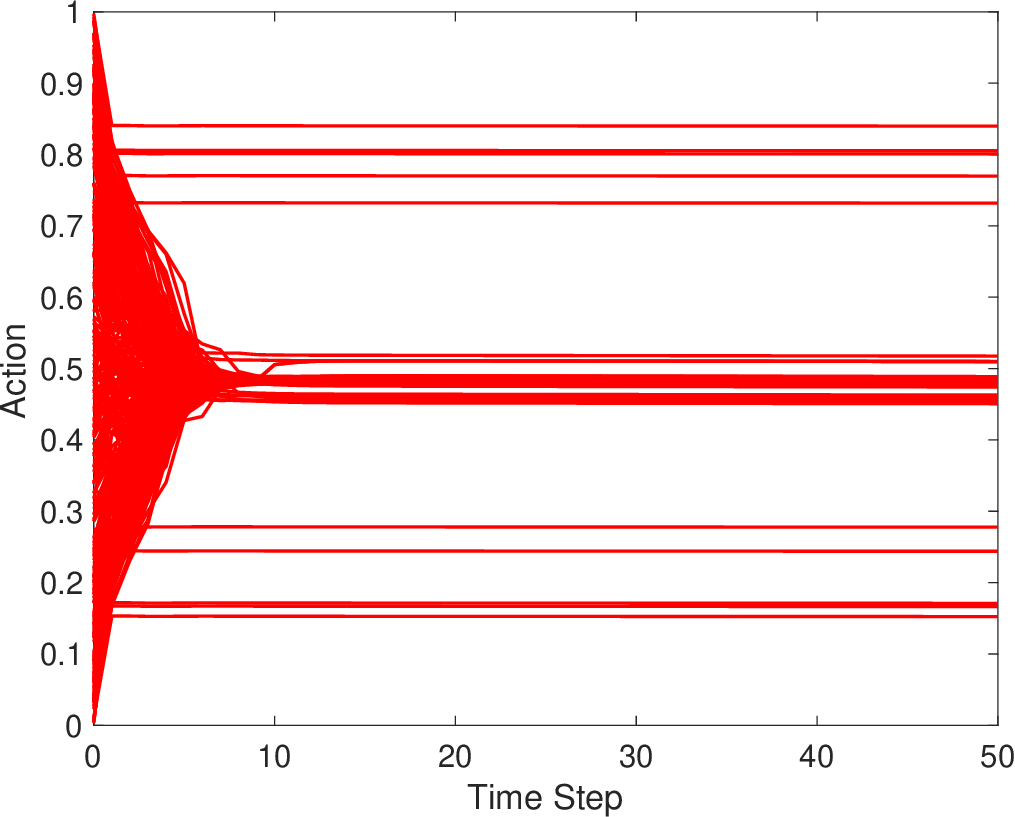}%
\label{fig8.2}}
\hfil
\subfloat[]{\includegraphics[width=2.0in]{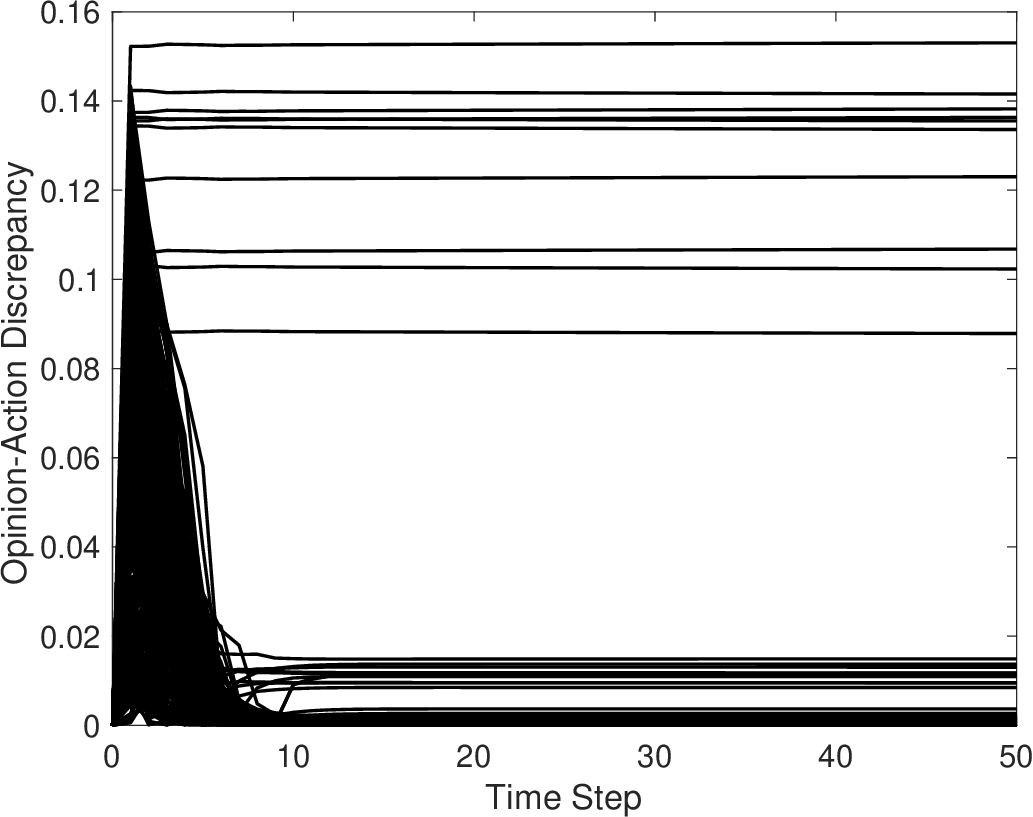}%
\label{fig8.3}}
\caption{Social dynamics for \(SW(300,6,0.8)\) under \(\epsilon=0.25\) and \(\phi=0.7\). (a) Opinion. (b) Action. (c) Opinion-action discrepancy.}
\label{fig8}
\end{figure*}

\begin{figure*}[!t]
\centering
\subfloat[]{\includegraphics[width=2.0in]{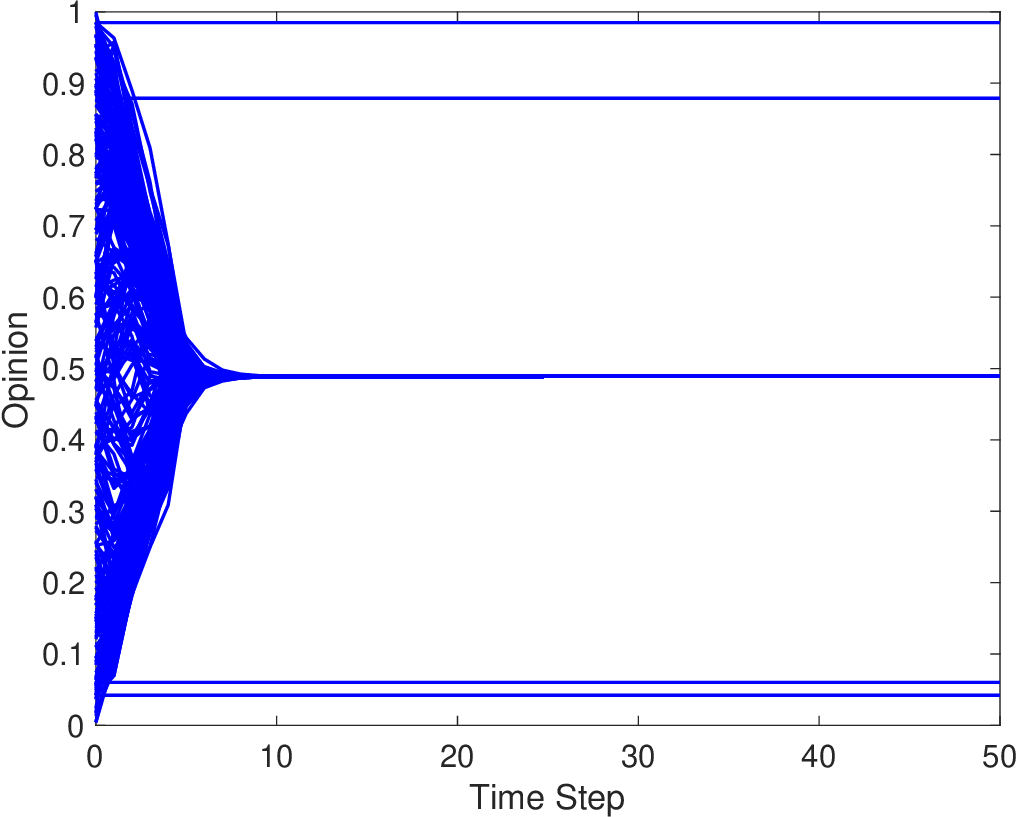}%
\label{fig9.1}}
\hfil
\subfloat[]{\includegraphics[width=2.0in]{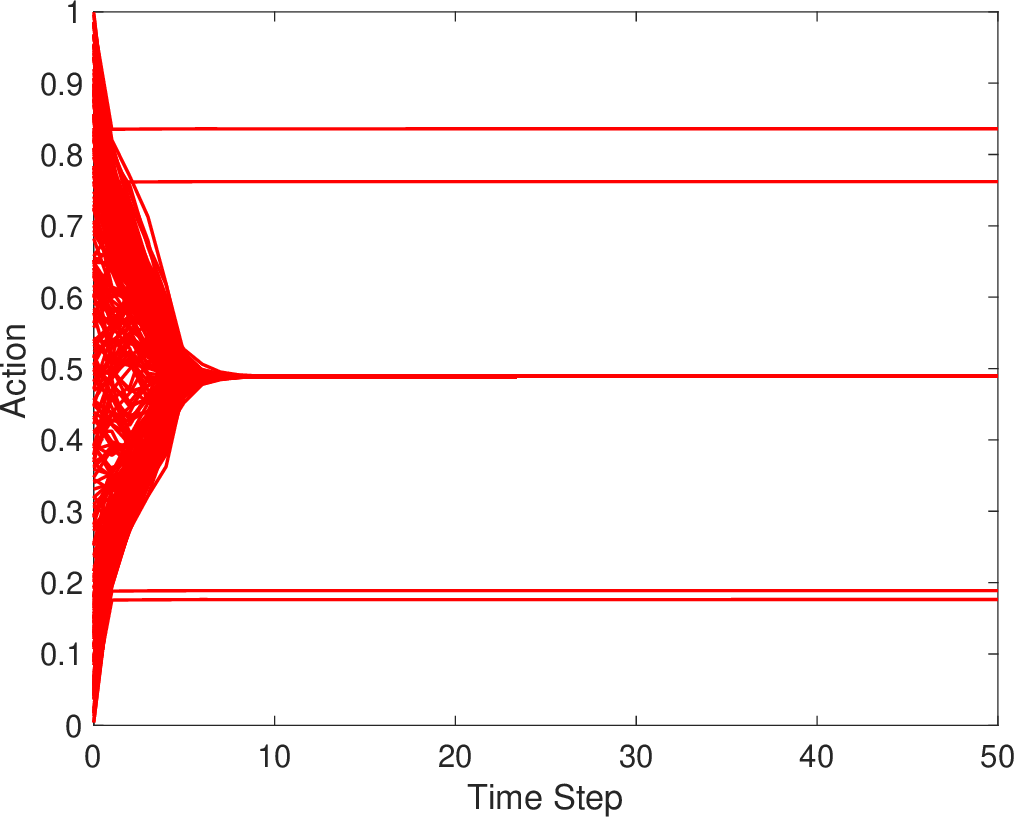}%
\label{fig9.2}}
\hfil
\subfloat[]{\includegraphics[width=2.0in]{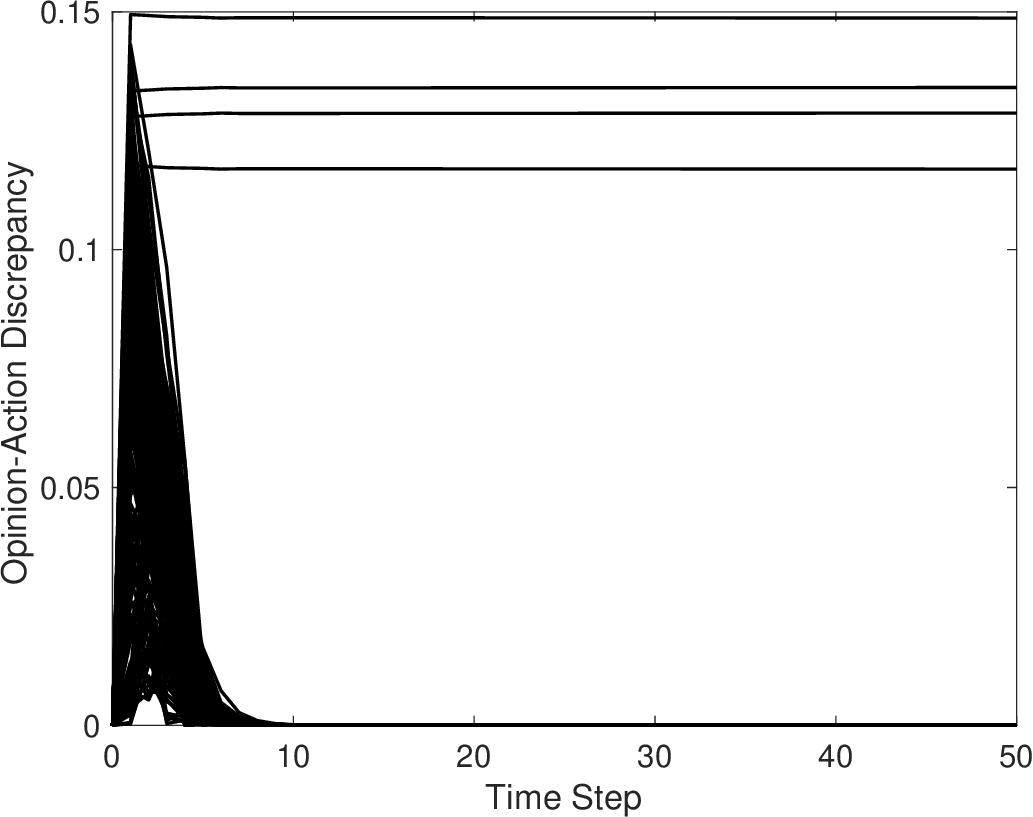}%
\label{fig9.3}}
\caption{Social dynamics for \(SF(300,9,6)\) under \(\epsilon=0.25\) and \(\phi=0.7\). (a) Opinion. (b) Action. (c) Opinion-action discrepancy.}
\label{fig9}
\end{figure*}

In this subsection, we examine the impact of social network topology on our model dynamics by incorporating two commonly studied network topologies introduced in Appendix~\ref{Appendix B}, namely the small-world network (SWN) and scale-free network (SFN). In both cases, the neighbor set of each agent is defined by an intersection of the bounded confidence principle and their underlying connections in the network. The small-world network is initialized with \(n=300\) nodes, an average node degree of \(k=6\), and a rewiring probability of \(p=0.8\), while the scale-free network is initialized with \(n=300\) nodes, \(m_0=9\) nodes in the initial complete graph, and \(m=6\) connections per newly added node. The openness and commitment parameters are set to \(\epsilon=0.25\) and \(\phi=0.7\), respectively, consistent with the setting of the last scenario in Section~\ref{section 4.1}. The results are presented in Fig.~\ref{fig8} and \ref{fig9}, respectively.

It is apparent from the results that the introduction of a social network disrupts opinion-action consistency within the group. Compared to the complete graph case as shown in Fig.~\ref{fig6}, where perfect alignment between collective opinion and action is achieved, both network topologies result in some agents' actions diverging from their opinions. Moreover, the presence of social network increases the number of opinion and action clusters at steady state, breaking the consensus achieved in Fig.~\ref{fig6}. This observation is consistent with the findings reported in Gir\'aldez-Cru et al. \cite{ref23}, which concluded that the existence of a social network impedes consensus formation and leads to the emergence of multiple intermediate opinion clusters. A comparison between Fig.~\ref{fig8} and Fig.~\ref{fig9} reveals that the small-world network exhibits more evident fragmentation than the scale-free network, which is also aligned with the conclusions drawn in \cite{ref23}. This phenomenon is driven by the prominent community structure and high local clustering inherent to the small-world network, which are found to inhibit the system from reaching global agreement \cite{ref24}.   

\subsection{Impact of Committed Minority in Social Diffusion} \label{section 4.4} 

As discussed earlier, this subsection introduces 20\% of committed innovators into the population to examine how a dedicated minority group shapes the social diffusion process under varying degrees of openness and commitment exhibited by non-stubborn individuals. The social diffusion process has been widely explored in experimental contexts, including studies by Centola and Baronchelli \cite{ref25}, Centola et al. \cite{ref26}, Amato et al. \cite{ref27}, and Ye et~al. \cite{ref28}. While these works provide valuable empirical insights, there has been a lack of theoretical modeling framework capable of systematically explaining the observed dynamics. Zino et al. \cite{ref13} addressed this gap by proposing a coevolution modeling framework that manages to capture phenomena such as the emergence of unpopular norms, popular disadvantageous norms, and paradigm shifts. However, all these studies, including Zino et al. \cite{ref13}, are constrained to binary decision-making scenarios, where each individual can only choose between two discrete options, denoted by \(-1\) and \(+1\). In contrast, each agent’s action is defined as a continuous variable in our model, as explained in Section~\ref{section 3}, which captures a smooth behavioral transition.

When the set of non-stubborn agents, i.e.,  \( i \in V \setminus S \), is assigned varied ranges of openness~(\(\epsilon_i\)) and commitment (\(\phi_i\)) parameters, we are able to observe and explain three typical outcomes of a general social diffusion process, namely, the adoption of innovation, the rejection of innovation, and the enforcement of unpopular norms, as reported in socio-psychological studies. 

\subsubsection{Adoption of Innovation} \label{section 4.4.1}

First, we hypothesize that when most non-stubborn agents are endowed with sufficiently high levels of openness (\(\epsilon_i\)) and commitment (\(\phi_i\)), they are more likely to adopt the innovative norm due to strong influence of subjective norms on both their opinions and actions. To validate this hypothesis, we initialize the openness and commitment of each non-stubborn agent based on the following uniform distributions: \(\epsilon_i \sim U[0.25, 0.3]\) and \(\phi_i \sim U[0.7, 0.8]\), respectively.

\begin{figure*}[!t]
\centering
\subfloat[]{\includegraphics[width=2.5in]{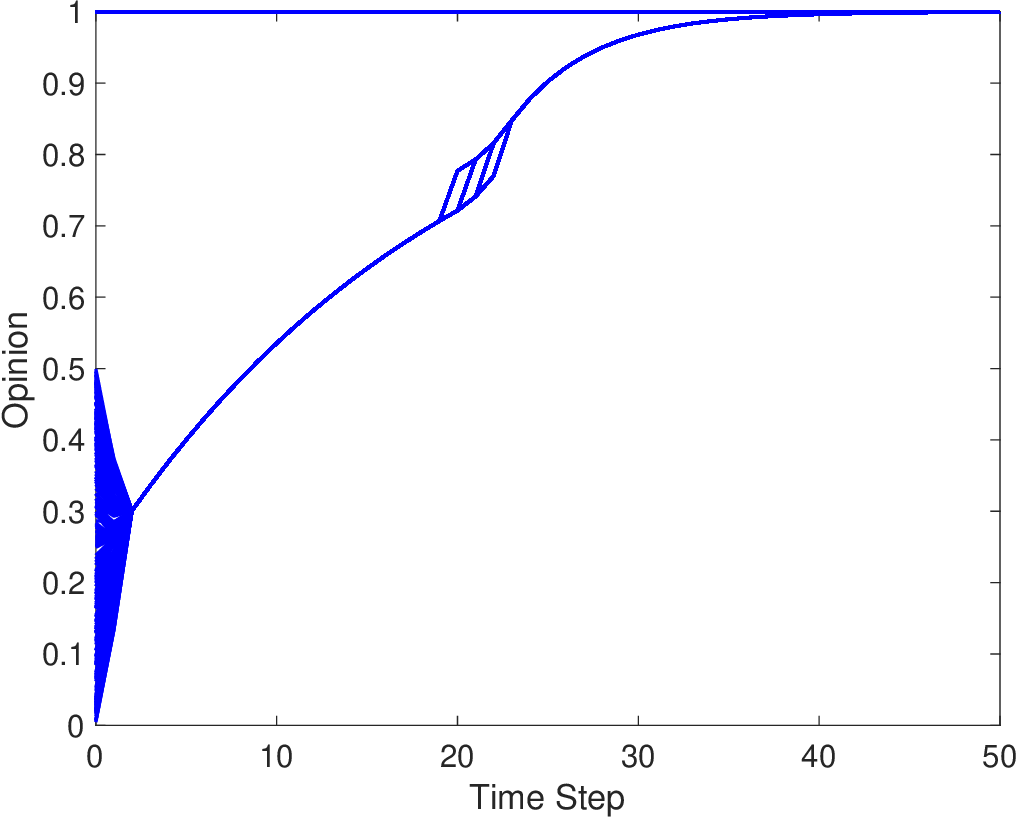}%
\label{fig10.1}}
\hfil
\subfloat[]{\includegraphics[width=2.5in]{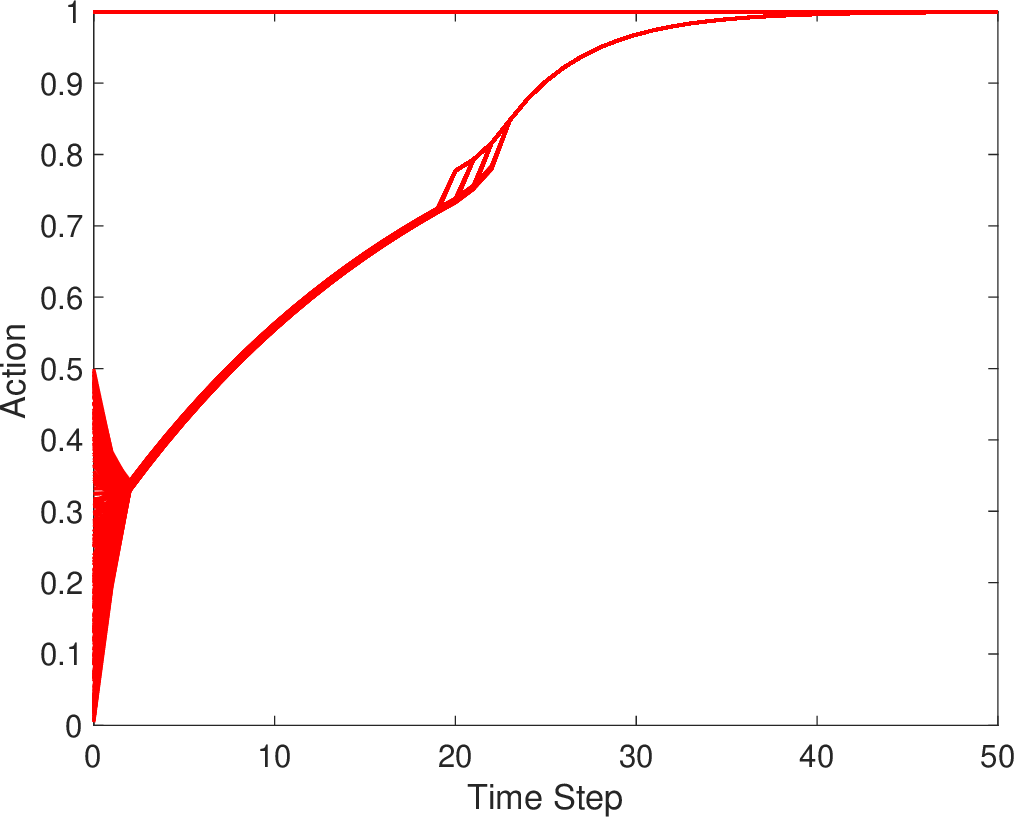}%
\label{fig10.2}}
\caption{Social dynamics for \(\epsilon_i \in [0.25,0.3]\) and \(\phi_i \in [0.7,0.8]\) (Adoption of innovation). (a) Opinion. (b) Action.}
\label{fig10}
\end{figure*}

As shown in Fig.~\ref{fig10}, after a brief transient period, all flexible agents' opinions and actions converge from the initial range \( [0,0.5] \) to the innovative norm, denoted by \(1\). In other words, the innovation is adopted by all agents and replaces the group's initial norm, a phenomenon also known as `paradigm shift' \cite{ref29} in the literature. This result indicates that when the majority of flexible agents are both open to diverse perspectives and strongly influenced by the group's perceived social norms, they are significantly more likely to accept and adopt a new idea, which facilitates the innovation's rapid diffusion throughout the population. For example, the rapid global adoption of Facebook, initially designed for U.S. college students, was likely driven by their high acceptance to innovation and strong peer pressure effects \cite{ref30}.    

\subsubsection{Rejection of Innovation} \label{section 4.4.2}

In contrast to Section~\ref{section 4.4.1}, we intuitively assume that when most flexible agents are close-minded, characterized by low \(\epsilon_i\), while strongly commit their actions to opinions, characterized by high \(\phi_i\), then they will have a tendency to maintain their initial norms, i.e., reject the innovation. In this case, the openness and commitment parameters of each non-stubborn agent are drawn from the following uniform distributions, with \(\epsilon_i \sim U[0.05, 0.1]\) and \(\phi_i \sim U[0.7, 0.8]\). 

\begin{figure*}[!t]
\centering
\subfloat[]{\includegraphics[width=2.5in]{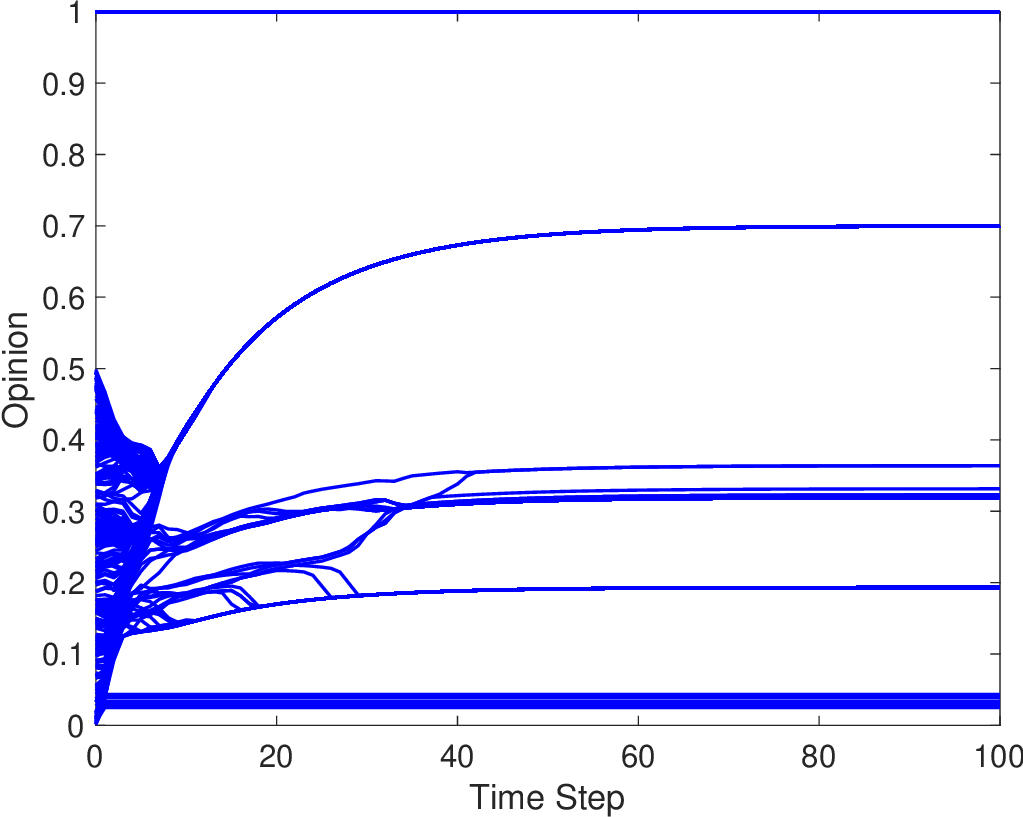}%
\label{fig11.1}}
\hfil
\subfloat[]{\includegraphics[width=2.5in]{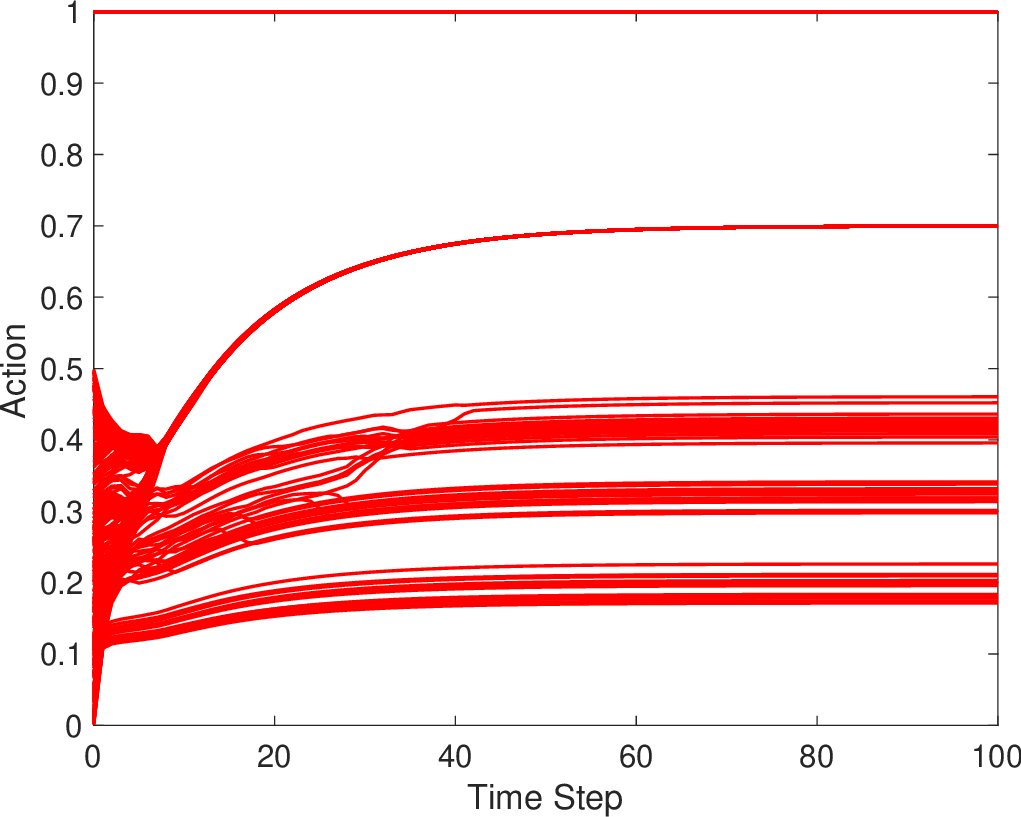}%
\label{fig11.2}}
\caption{Social dynamics for \(\epsilon_i \in [0.05,0.1]\) and \(\phi_i \in [0.7,0.8]\) (Rejection of innovation). (a) Opinion. (b) Action.}
\label{fig11}
\end{figure*}

The results depicted in Fig.~\ref{fig11} confirm our assumption since most flexible agents' opinions and actions remain in the initial interval \([0, 0.5]\). Simiarly, our hypothesis is also supported by the findings of a well-known socio-psychological case study. In \cite{ref31}, health workers persuaded housewives in a Peruvian town to boil their drinking water as a measure to improve the local hygiene. However, the sanitation campaign failed to induce notable behavioral change among residents, which was largely attributed to a deeply entrenched local custom that associated boiling water with illness. These results indicate that deeply rooted social norms could serve as a formidable barrier to the diffusion of new practices.   

\begin{figure*}[!t]
\centering
\subfloat[]{\includegraphics[width=2.5in]{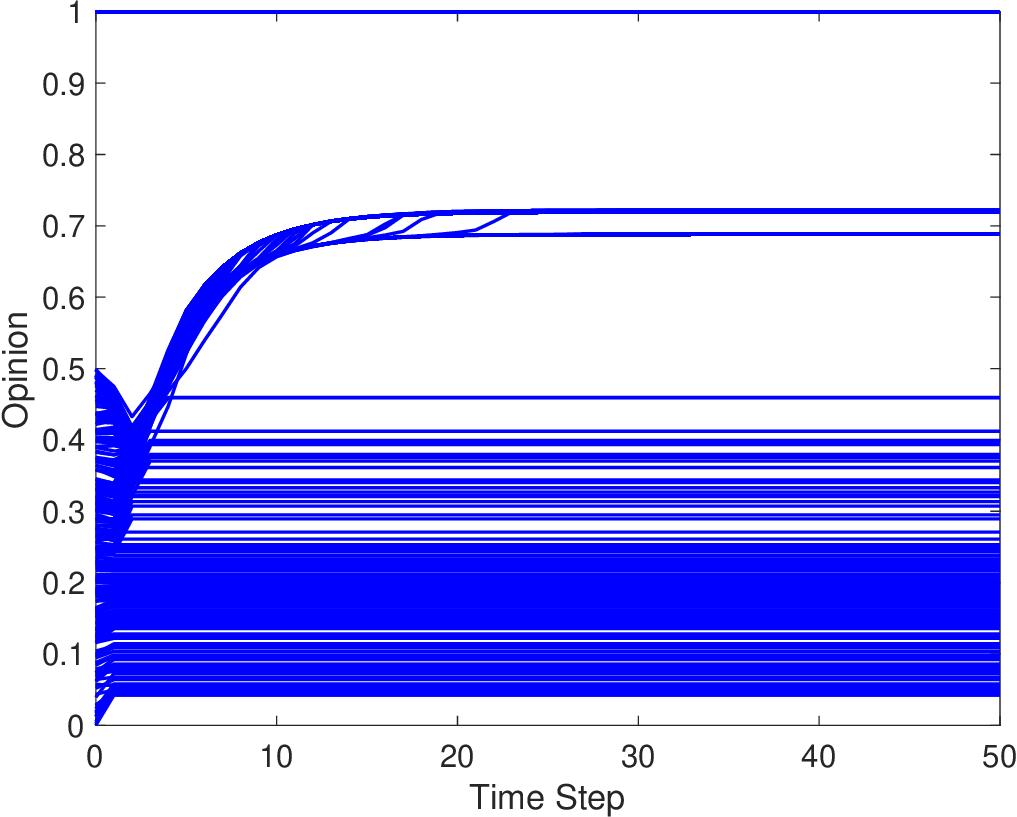}%
\label{fig12.1}}
\hfil
\subfloat[]{\includegraphics[width=2.5in]{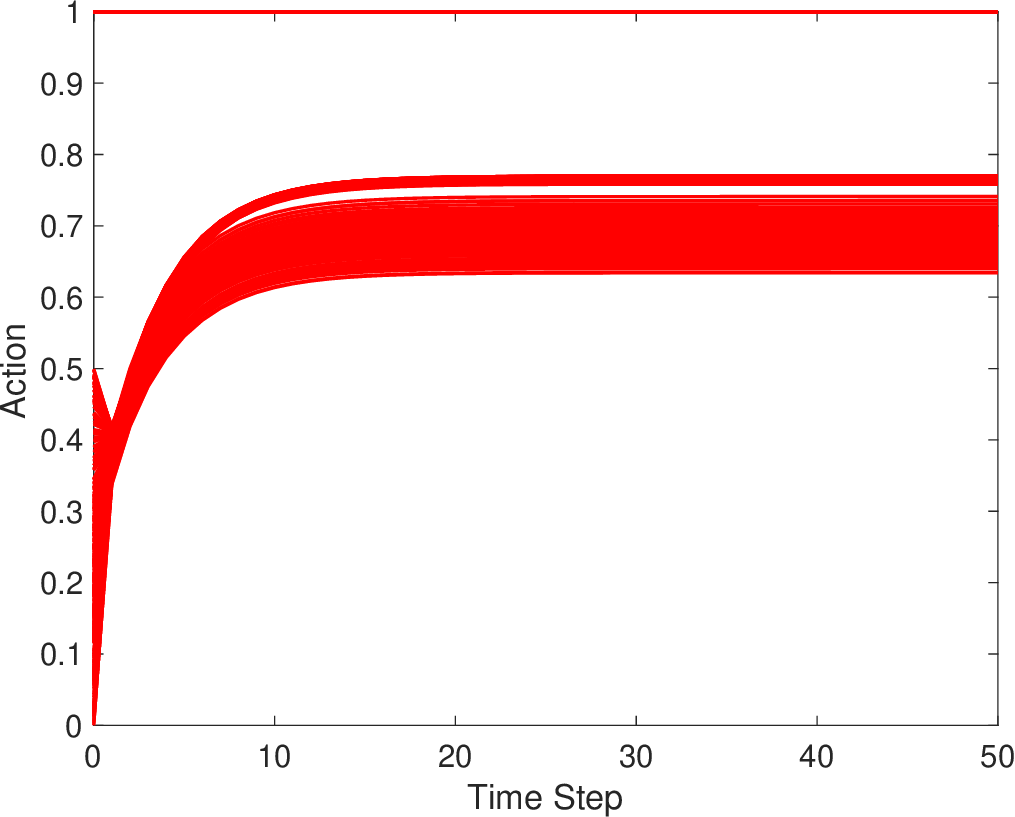}%
\label{fig12.2}}
\caption{Social dynamics for \(\epsilon_i \in [0.05,0.1]\) and \(\phi_i \in [0.1,0.2]\) (Enforcement of unpopular norms). (a) Opinion. (b) Action.}
\label{fig12}
\end{figure*}

\subsubsection{Enforcement of Unpopular Norms} \label{section 4.4.3}

Finally, we observe a distinct behavioral pattern when flexible agents are closed to differing opinions (low \(\epsilon_i\)) but their actions are strongly influenced by subjective norms (low \(\phi_i\)). In this case, the majority's actions are driven toward the innovative norm, yet their opinions remain largely unchanged. The results are shown in Fig.~\ref{fig12}, where each flexible agent's openness and commitment are initialized following: \(\epsilon_i \sim U[0.05, 0.1]\) and \(\phi_i \sim U[0.1, 0.2]\), respectively. It is apparent that the agents collectively enforce an action closer to the innovative norm, within the interval \([0.6, 0.75]\), while their opinions are mostly distributed in the initial interval \([0, 0.5]\). In other words, most agents enforce a norm that they do not genuinely endorse, referred to as `enforcement of unpopular norms' \cite{ref17}.  

There are many empirical cases in which individuals are compelled to publicly support behavior they privately oppose \cite{ref9,ref10,ref11}. As summarized in \cite{ref17}, the enforcement of unpopular norms shares the same underlying mechanism as pluralistic ignorance. The key drivers for such phenomena include the pressure to conform to subjective norms and the misleading belief that others' similar behavior genuinely reflect their private attitudes \cite{ref11}. For example, in the context of alcohol consumption on campus, students' private attitudes toward drinking are uniformly distributed across a scale from 0 to 10, yet their estimates of others' average comfort with drinking behavior follow a normal distribution centered at around 7 \cite{ref11}. These findings are consistent with our observed results, and reveal the role that normative pressure and misperceptions of collective actions play in the formation of unpopular norms. 

\section{Implications for Human Social Systems} \label{section 5}
While most existing socio-psychological case studies lack the requisite data for a direct validation of our model, our findings provide valuable insights that could guide the design of future experiments. By refining data collection methodologies to align with the key parameters of our model, socio-psychological case studies could offer a robust framework for assessing the model's predictive capabilities in real-world social contexts. 

Given that our model incorporates two key parameters---namely, \(\epsilon\) (openness) and \(\phi\) (commitment)---that reflect individual personality traits, it is imperative that future socio-psychological case studies collect empirical data on these traits from participants in order to facilitate partial validation of our model. To effectively capture an individual's openness and commitment traits, it is essential to employ a survey-based approach, which requires careful design of survey questions and a well-defined response scale (e.g., a 5-point Likert scale) to obtain quantifiable and meaningful data. For example, an appropriate question to measure one's \textit{openness} level could be: \textit{``When making decisions on a controversial issue, how likely are you to adjust your opinion if you observe that others disagree with your prespective?''} The scale for responses could be designed to range from 1 to 5, where 1 represents ``not likely at all'' and 5 represents ``extremely likely''. Similarly, a suitable question to assess one's \textit{commitment} level could be: \textit{``When making decisions on a controversial issue, how likely are you to stick to your own opinion even when faced with strong social pressure to conform to the majority?''} A higher response value indicates a greater level of openness or commitment demonstrated by the participant. Researchers can then analyze the distribution of responses across all participants to assess the statistical properties of these traits within the population.    

The various socio-psychological scenarios discussed earlier can be broadly classified into three categories based on the relationship between individuals' opinions and actions: (1) opinion-action divergence, where most individuals hold private opinions that differ significantly from their public actions; (2) fragmented distribution, where individuals' opinions and actions are mostly aligned but the population remains divided into multiple clusters; and (3) consensus formation, where almost all individuals' opinions and actions converge to the same single shared value.  

To illustrate how experimental data could partially validate our model, we consider the first category of phenomenon---opinion-action divergence---by taking Prentice and Miller \cite{ref11} as an example. In \cite{ref11}, it has been reported that there is a clear distinction between students' private attitudes on alcohol consumption and their displayed acceptance. If this case study had been conducted with the addition of a survey measuring each student's personality traits, the collected data would have offered empirical insights for our model. Specifically, if the measured openness (\(\epsilon\)) and commitment (\(\phi\)) among all participants exhibited a distribution skewed toward lower values---indicating that most students are resistant to changing their opinions and do not align their actions with their opinions, this would correspond to our model's 	parameter setting for opinion-action divergence. Such consistency between empirical data and theoretical model settings would provide evidence that our model effectively captures the underlying mechanisms of this social phenomenon and has the potential for prediction in similar application contexts.

Simiarly, suppose health workers in \cite{ref31} had collected survey responses from Peruvian housewives and the results revealed that most participants exhibited low openness values and high commitment values, this would be consistent with our model's parameter setting and would offer solid evidence supporting the predictive capability of our model in other related social systems. Finally, the yielding subjects in Asch's experiment \cite{ref10} actually reported experiencing overwhelming confusion during the decision-making process and were easily swayed by the unanimous opinion of others, which suggests a high level of openness and therefore aligns with our model's parameter setting. In conclusion, our model offers critical insights for designing future socio-psychological experiments, which can in turn provide strong empirical support for its predictive validity across different social contexts.

\section{Conclusions and future work} \label{section 6}
In this paper, we proposed a novel modeling framework that integrates opinion dynamics with a decision-making mechanism to delve into the interplay between individuals' opinions and actions. Specifically, we generalized the classical Hegselmann-Krause (HK) model, a well-known instance of bounded confidence opinion dynamics (BCOD), by combining it with a utility maximization problem. To the best of our knowledge, this is the first work that utilizes an agent-based BCOD model in conjunction with the optimization approach to develop a coevolution modeling framework between opinions and actions. Compared to the few existing coevolution models, particularly the one proposed by Zino et al. \cite{ref13}, our model introduces several key improvements, including: (1) it extends the action/decision variable range from binary choices to a continuous domain, broadening our model's applicability to encompass a more general context; (2) it explicitly distinguishes between one's private opinion and public action based on their visibility properties, aligning the model more closely with real-world scenarios; and (3) it introduces the concept of subjective norms into agents' utility functions, inspired by the Theory of Planned Behavior (TPB), to provide a more socio-psychologically grounded framework.       

Simulation results from our model demonstrate that the magnitude of opinion-action divergence within a group can be controlled by adjusting two key parameters: \(\epsilon\) and \(\phi\), which characterize each agent's openness and commitment, respectively. In particular, when \(\epsilon\) and \(\phi\) are large enough such that their linear combination \(10\epsilon + 3\phi\) exceeds a threshold of 3.5, all agents' opinions align with their actions, resulting in complete consensus. Conversely, when \(\epsilon\) is small but \(\phi\) is large, the population splits into multiple clusters of opinions and actions. In cases where both \(\epsilon\) and \(\phi\) are very small, significant divergence between agents' opinions and actions emerge, possibly leading to pluralistic ignorance. Our findings also reveal that the presence of a social network impedes consensus formation and increases the average opinion-action discrepancy within the population. Finally, by introducing a small number of committed agents into the group, we utlized our theoretical framework to investigate the social diffusion process and identified three typical phenomena: adoption of innovation, rejection of innovation, and enforcement of unpopular norms. These observed outcomes closely correspond to findings in socio-psychological studies, providing strong support for the underlying mechanism of our model and demonstrating its conceptual validity. 

Contemporary societal and scientific challenges underscore the necessity for a deeper understanding of the intricate interactions among cyber, physical, and human-social systems. A critical gap in bridging these systems lies in the absence of dynamic modeling frameworks capable of effectively capturing and representing human-social behavior. Although significant achievements have been made in behavioral economics, the Theory of Planned Behavior (TPB), and opinion dynamics, the question of how to accurately predict behavioral changes, whether at the individual or aggregate level, remains a formidable research task. For future work, we will focus on integrating the elements of subjective control, intent, and uncertainties inherent in human behavior, as outlined in TPB. Our ultimate goal is to develop a robust and realistic model of human-social behavior, which can be utilized for predictive purposes within the context of cyber-physical-human systems.

%\section*{Acknowledgments}
%This should be a simple paragraph before the References to thank those individuals and institutions who have supported your %work on this article.

%{\appendix[Proof of the Zonklar Equations]
%Use $\backslash${\tt{appendix}} if you have a single appendix:
%%Do not use $\backslash${\tt{section}} anymore after $\backslash${\tt{appendix}}, only $\backslash${\tt{section*}}.
%If you have multiple appendixes use $\backslash${\tt{appendices}} then use $\backslash${\tt{section}} to start each appendix.
%You must declare a $\backslash${\tt{section}} before using any $\backslash${\tt{subsection}} or using $\backslash${\tt{label}} ($\backslash${\tt{appendices}} by itself
%starts a section numbered zero.)}

{\appendices
\section{Comparison of the DW and HK models} \label{Appendix A}
In the DW model, the opinion update process is based on a random pairwise interaction mechanism \cite{ref5}. At each time step, one pair of agents is randomly selected to interact with each other. If their opinions fall within each other's confidence area, they will adjust their opinions towards the average. This mechanism implies that at most two agents' opinions are updated during one iteration. Thus, the DW model is especially suitable for modeling pairwise interactions, such as gossip exchanges between two individuals \cite{ref32}.

The Hegselmann-Krause (HK) model, on the other hand, operates on a parallel updating mechanism \cite{ref6}. At each time step, all agents simultaneously update their opinions by identifying their neighbors---agents whose opinions fall within their confidence area---and averaging these neighbors' opinions. This synchronous updating process enables the HK model to effectively represent large-scale interactions, e.g., formal meetings where experts discuss specific issues \cite{ref32}.

\section{Common social network topologies} \label{Appendix B}
A complete graph, denoted as \( K_n \), is a simple undirected graph in which every pair of distinct vertices is connected by a unique edge \cite{ref18}. In other words, a complete graph \( K_n \) is defined as a graph \( G(V, A) \) consisting of \( n \) nodes, where \( n = |V| \), and \( A_{ij} = 1 \) for every pair of nodes \( i, j \in V \) such that \( i \neq j \). The fully connected structure facilitates the study of opinion dynamics in well-mixed systems, where agents interact without restrictions. Most classical BCOD models, including the DW and HK models, assume a complete graph in their initial formulations. 

The small-world network, introduced by Watts and Strogatz \cite{ref19}, is constructed by introducing randomness into a regular lattice through the rewiring of edges with a probability \( p \). This process enables small-world networks to combine the high clustering property of regular lattices with the short characteristic path length of random graphs, a feature commonly referred to as six degrees of separation \cite{ref33}. The Watts-Strogatz model \cite{ref19}, used to generate a small-world network, is defined by three key parameters: \( n \), the total number of nodes in the network; \( k \), the average degree of each node; and \( p \), the rewiring probability of edges. For simiplicity, we refer to small-world networks as \( SW(n, k, p) \) throughout this paper.

The scale-free network, proposed by Barabási and Albert \cite{ref20}, is generated through a formation process governed by the preferential attachment mechanism. The process begins with a complete graph containing a small number of initial nodes, after which new nodes are steadily added to the network. The probability of each newly added node connecting to an existing node is proportional to the latter's degree, implying that newly added nodes prefrentially connect to existing nodes with higher degrees. As a result of preferential attachment mechanism, scale-free networks exhibit a power-law degree distribution, where a small number of nodes possess high degrees, often referred to as hubs, while most nodes have fewer connections. The Barabási-Albert model \cite{ref20} for generating a scale-free network is also characterized by three parameters: \( n \), the total number of nodes in the network; \( m_0 \), the number of nodes in the initial complete graph; and \( m \), the number of existing nodes connected to each newly added node, where \( m \leq m_0 \). For ease of reference, scale-free networks are denoted as \( SF(n, m_0, m) \) in this paper.
}

%\newpage

%\begin{IEEEbiography}[{\includegraphics[width=1in,height=1.25in,clip,keepaspectratio]{fig1}}]{Michael Shell}
%Use $\backslash${\tt{begin\{IEEEbiography\}}} and then for the 1st argument use $\backslash${\tt{includegraphics}} to declare and link the author photo.
%Use the author name as the 3rd argument followed by the biography text.
%\end{IEEEbiography}

\vspace{100pt}

\begin{IEEEbiographynophoto}{Chen Song}
Chen Song received the B.Eng. degree in Automation from South China University of Technology, Guangzhou, China, in 2020, and the M.Sc. degree in Computer Control and Automation from Nanyang Technological University, Singapore, in 2021, respectively. Currently, he is a Ph.D. student in the School of Electrical and Electronic Engineering, Nanyang Technological University, Singapore. His research interests include computational social systems, cyber-physical-human systems, and social psychology.
\end{IEEEbiographynophoto}

\vspace{-350pt}

\begin{IEEEbiographynophoto}{Vladimir Cvetkovic}
Vladimir Cvetkovic received the B.Sc. and M.Sc. degrees in civil engineering from the University of Belgrade, Belgrade, Serbia, in 1978 and 1980, respectively, and the Ph.D. degree in civil engineering from KTH Royal Institute of Technology, Stockholm, Sweden, in 1986.,His research has mostly focused on safety and risk assessment in complex technical systems related, in particular, to critical infrastructures. He has also worked in geoengineering of rock fracture networks with application to mass transport under uncertainty. More recently his research addresses integration of infrastructures (such as water, energy, waste, and transportation) in urban systems.
\end{IEEEbiographynophoto}

\vspace{-350pt}

\begin{IEEEbiographynophoto}{Rong Su}
Rong Su (Senior Member, IEEE) received the B.Eng. degree in Automation from the University of Science and Technology of China, Hefei, China, in 1997, and the M.A.Sc. and Ph.D. degrees in Electrical and Computer Engineering from the University of Toronto, in 2000 and 2004, respectively. Currently, he is a full Professor in the School of Electrical and Electronic Engineering, Nanyang Technological University, Singapore. He was affiliated with the University of Waterloo and Technical University of Eindhoven before he joined Nanyang Technological University in 2010. His research interests include multi-agent systems, cybersecurity of discrete-event systems, supervisory control, model-based fault diagnosis, control and optimization in complex networked systems with applications in flexible manufacturing, intelligent transportation, human–robot interface, power management and green buildings. He has authored or coauthored over 230 journal and conference publications, one monograph, and nine granted/filed patents. Dr. Su is an Associate Editor for Automatica, Journal of Discrete Event Dynamic Systems: Theory and Applications, and Journal of Control and Decision. He was the Chair of the Technical Committee on Smart Cities in the IEEE Control Systems Society from 2016 to 2019, and is currently a Co-Chair of IEEE Robotics and Automation Society Technical Committee on Automation in Logistics, and the Chair of IEEE Control Systems Chapter, Singapore.
\end{IEEEbiographynophoto}

\end{document}